\documentclass[aps,prm, reprint,superscriptaddress, numerical, amsmath, amssymb]{revtex4-1}

\usepackage{graphicx,epsfig,float}
\usepackage[varg]{txfonts}

\allowdisplaybreaks
\begin{document}
	\title{Magnetic and electronic properties of 1D hybrid nanoobjects composed of alternating polycyclic hydrocarbon regions and double carbon chains}
		
	\author{Irina V. Lebedeva}
	\email{liv\_ira@hotmail.com}
	\affiliation{CIC nanoGUNE BRTA, San Sebasti\'an 20018, Spain}
	\affiliation{Simune Atomistics, Avenida de Tolosa 76, San Sebastian 20018, Spain}
	
	\author{Sergey A. Vyrko}
	\email{vyrko@bsu.by}
	\affiliation{Physics Department, Belarusian State University, Nezavisimosti Ave.~4, Minsk 220030, Belarus}
	
	\author{Alexander S. Sinitsa}
	\email{alexsinitsa91@gmail.com}
	\affiliation{National Research Centre ‘‘Kurchatov Institute’’, Kurchatov Square~1, Moscow 123182, Russia}
	\affiliation{Kintech Lab Ltd., 3rd~Khoroshevskaya Street~12, Moscow 123298, Russia}
	
	\author{Sergey V. Ratkevich}
	\email{ratkevich@bsu.by}
	\affiliation{Physics Department, Belarusian State University, Nezavisimosti Ave.~4, Minsk 220030, Belarus}
	
	\author{Andrey M. Popov}
	\email{popov-isan@mail.ru}
	\affiliation{Institute of Spectroscopy of Russian Academy of Sciences, Fizicheskaya str.~5, Troitsk, Moscow 108840, Russia}
		
	\author{Andrey A. Knizhnik}
	\email{knizhnik@kintechlab.com}
	\affiliation{National Research Centre ‘‘Kurchatov Institute’’, Kurchatov Square~1, Moscow 123182, Russia}
	\affiliation{Kintech Lab Ltd., 3rd~Khoroshevskaya Street~12, Moscow 123298, Russia}
	
	\author{Nikolai A. Poklonski}
	\email{poklonski@bsu.by}
	\affiliation{Physics Department, Belarusian State University, Nezavisimosti Ave.~4, Minsk 220030, Belarus}
	
	\author{Yurii E. Lozovik}
	\affiliation{Institute of Spectroscopy of Russian Academy of Sciences, Fizicheskaya str.~5, Troitsk, Moscow 108840, Russia}

\begin{abstract}
	It has been proposed recently that 1D hybrid nanoobjects consisting of alternating double carbon chains and polycyclic carbon regions can be obtained from graphene nanoribbons of alternating width by electron irradiation. Here, based on density functional theory calculations, we show that magnetic and electronic properties of such nanoobjects can be changed dramatically by modifying the chain length and edge structure of polycyclic regions and this opens wide possibilities for spintronic applications. Nanoobjects composed of polycyclic regions with dangling bonds and even chains are found to behave as magnetic semiconductors that can generate spin-polarized currents. Band gaps of nanoobjects with odd chains change considerably upon switching between magnetic states making them promising for magnetic tunnel junctions. We also demonstrate that use of a hybrid exchange-correlation functional is important to properly describe stability of magnetic states, band gaps and synergistic effects of nanoobject components leading, for example, to magnetism in even chains.
\end{abstract}

\maketitle


\section{Introduction}
Because of the unique electronic properties, graphene is considered as a basis for future nanoelectronics \cite{Geim2009, Geim2009a, Sato2015, Westervelt2008}. One of the main advantages of graphene is the possibility to tune its properties in wide ranges by manipulation of the atomic structure: cutting nanostructures of a particular shape, modifying edges, creating atomic defects, etc. \cite{Banhart2011, Terrones2012, Skowron2015, Bhatt2022} The band gap of graphene nanoribbons (GNRs) can be controlled by the edge type and termination, GNR width and presence of local defects \cite{Han2007, Ritter2009, Son2006, Fujita1996, Kobayashi2005, Niimi2006, Tao2011, Koskinen2008, Li2010, Gunlycke2007, Bhandary2010, Lee2009, Kunstmann2011, Nakada1996, Lebedeva2012a, Cantele2009, Son2006a, Hod2007}. Furthermore, spin ordering \cite{Son2006, Kunstmann2011, Seitsonen2010, Magda2014, Cheng2012, Cheng2012prb, Gan2010, Bhandary2010, Lee2009, Song2010, Wang2016, Brede2023, Palacios2010} is observed at edges of zigzag GNRs and the possibility to use GNRs for spintronic applications has been actively explored in the recent years \cite{Zhang2014, Zeng2011, Magda2014, Song2010, Zhang2021a, Kang2017, Rezapour2020, Han2014, Wimmer2008, Yazyev2008, Yazyev2010, Son2006a, MunozRojas2009, Kim2008, Zhang2017, Cantele2009, Soriano2010}. Giant magnetoresistance has been predicted for GNRs because of switching between different magnetic states under the magnetic field applied \cite{MunozRojas2009, Kim2008}. It has been also proposed to induce half-metallicity in GNRs, i.e. make them behave as a metal for one spin component and an insulator for the other, by application of an electric field across the GNR \cite{Son2006a}, GNR bending \cite{Zhang2017}, defect creation \cite{Cantele2009} or functionalization \cite{Hod2007}. Other carbon nanostructures also hold promise for electronic applications. Carbon atomic chains, for example, represent the finest possible wires \cite{Ravagnan2009, Poklonski2013, Zanolli2010, Liu2013, Artyukhov2014, LaTorre2015, Borrnert2010, Crljen2007, Lang1998, Zeng2010, Zhou2017, Liang2019, Cretu2013, Chen2009, Liu2015, Cahangirov2010}. The transition from a conductive to insulating state \cite{Ravagnan2009, Zanolli2010, Liu2013, Artyukhov2014, LaTorre2015} is observed upon changing the type of bond network in the chain.

Combining carbon structures of different types makes possible to achieve intriguing properties via the synergistic effects of the components \cite{Poklonski2019}. Therefore, significant efforts are made to propose and synthesize new hybrid nanoobjects. The combination of one-dimensional carbon nanotubes and two-dimensional graphene materials to generate three-dimensional carbon nanotube–graphene hybrid thin films is actively studied for transparent conductors, electron field emitters, field-effect transistors, biosensors, supercapacitors, hydrogen storage and other diverse applications \cite{Xia2017, Dang2016, Vinayan2012}. Carbon films combining different allotropic forms of carbon are considered for biosensors \cite{Laurila2017}. The combination of carbon nanotubes and nanofibes has been proposed for environmental applications \cite{Navrotskaya2020}. Two-dimensional hybrid $sp-sp^2$ carbon systems such as graphyne and graphdiyne hold great promise for nanoelectronics, energy storage, membranes for gas separation, etc. \cite{Baughman1987, Casari2016, Srinivasu2012, Ivanovskii2013, Li2014}. 

\begin{figure}
	\centering
	\includegraphics[width=\columnwidth]{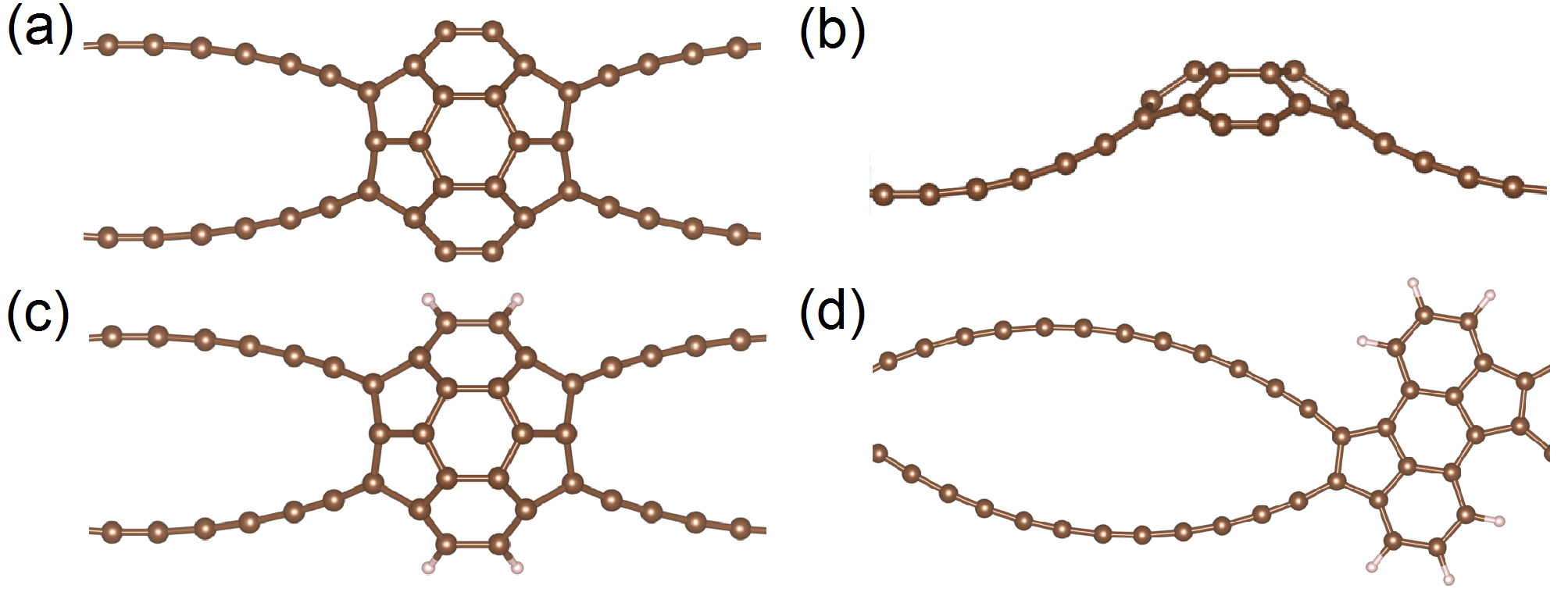}
	\caption{1D nanoobjects with alternating polycyclic regions and double carbon chains optimized using the PBE functional: (a) F-10, (b) F-10 (side view), (c) F-10H, (d) D1-12H. Carbon and hydrogen atoms are colored in brown and white, respectively. One unit cell is shown.
	}
	\label{fig:geom}
\end{figure}

Transmission electron microscopy (TEM) is a powerful tool to manipulate the structure of carbon materials in a controllable way \cite{Song2011, Sinitsa2023} and to transform and create nanoobjects \cite{Matsokin2023, Chuvilin2009, Chuvilin2010, Sloan2000, Sinitsa2018, Sinitsa2017, Jin2009}. It has been shown recently that hydrogen loss in parts of carbon nanostructures resembling narrow GNRs can lead to formation of double carbon chains \cite{Sinitsa2021, Tanuma2022, Matsokin2023}. In particular, as demonstrated by molecular dynamics simulations \cite{Sinitsa2021}, electron irradiation of GNRs of alternating width resulting in hydrogen removal from the GNR edges could provide a way to synthesize one-dimensional (1D) hybrid structures consisting of polycyclic regions alternating with double carbon chains (similar to ones shown in Fig. \ref{fig:geom}). Although such a process has not been yet realized experimentally, recent advances in synthesis of GNRs with precise atomic structure \cite{Cai2010, Brede2023, Zhou2020} and controlled transformation of nanostructures in TEM \cite{Chuvilin2009, Chuvilin2010, Sloan2000, Sinitsa2017, Jin2009} make us believe that it will be possible in the nearest future. Double carbon chains \cite{Chuvilin2009, Tanuma2022, Jin2009} and polyaromatic hydrocarbons with pentagons \cite{Lieske2023, Martin2019} have been observed experimentally suggesting that hybrid nanoobjects composed of them should be stable under experimental conditions. It has been demonstrated in many occasions that predictive modeling can anticipate experimental results. A famous example is graphyne that was studied theoretically since 1987 \cite{Baughman1987} and finally synthesized in 2022 \cite{Desyatkin2022, He2023}. A diamond-like layer (diamane) was proposed theoretically \cite{Chernozatonskii2009} a decade before its experimental discovery \cite{Bakharev2020}. Domain walls separating commensurate domains in few-layer graphene were predicted in 2011 \cite{Popov2011} and soon observed experimentally \cite{Alden2013}. In addition to predictions on structure and stability of new materials, theoretical methods provide a guidance on their possible applications \cite{Li2023, Davies2021, Gopalan2022, Dias2021}, i. e. serve to screen promising materials that deserve experimental realization. For this reason, in the present paper, we investigate the structure, magnetic and electronic properties of hybrid nanoobjects proposed in \cite{Sinitsa2021}. These properties originate from the interplay of two components of hybrid nanoobjects: polycyclic regions and carbon atomic chains. 

Molecular dynamics simulations \cite{Sinitsa2021} showed that different polycylic regions can be produced by GNR irradiation. According to the density functional theory (DFT) calculations performed in the same paper, nanoobjects with the polycyclic region F (Figs. \ref{fig:geom}a-c) are the most stable energetically. Electron irradiation cleans the nanostructure edges from hydrogen and other adsorbates \cite{Sinitsa2021, Tanuma2022, Matsokin2023} and such pristine edges are prone to formation of pentagon rings that help to reduce the edge length and stabilize dangling bonds. Formation of the polycyclic region F that resembles a fullerene cap in an irradiated GNR is thus driven by the same forces as exothermic processes of spontaneous transformation of graphene flakes to fullerenes \cite{Chuvilin2010, Lebedeva2008, Skowron2013}, formation of curved polyaromatic hydrocarbons, such as corannulene, in soot \cite{Lieske2023, Martin2019} and zigzag edge reconstruction \cite{Koskinen2008, Cheng2012prb, Cheng2012, He2015, Polynskaya2022, Polynskaya2022a}. It should be noted that in spite of the F region being the most energetically favourable, the highest yield of hybrid nanoobjects in the molecular dynamics simulations \cite{Sinitsa2021} corresponds not to the F region but to the D1 region with fewer pentagons (Fig. \ref{fig:geom}d) for kinetic reasons. There are barriers for transformation of the D1 region to other structures warranting its meta-stability \cite{Sinitsa2021}. Therefore, to investigate the effects of the structure of polycyclic region edges on the performance of hybrid nanoobjects, we consider both the F (Figs. \ref{fig:geom}a-c) and D1 polycylic regions (Fig. \ref{fig:geom}d). 

As mentioned above, edges of polycyclic regions are expected to be pristine, i.e. free of any adsorbates, during creation of hybrid nanoobjects by electron irradiation \cite{Sinitsa2021}. However, such pristine edges are usually rather reactive and easily get functionalized, e.g., by hydrogen \cite{Koskinen2008, Wassmann2008, Kunstmann2011, Bhandary2010}. Therefore, we also consider the influence of hydrogen termination on hybrid nanoobject properties.

Two types of bonding patterns are possible for carbon atomic chains: consecutive double bonds $=$C$=$C$=$ to give cumulenes \cite{Tommasini2014, Gu2008, Hino2003} and alternating single and triple bonds $-$C$\equiv$C$-$ to give polyynes \cite{Eisler2005, Yueze2022, Agarwal2016, Tommasini2014}. The $\pi$-system of carbon atomic chains has two valence electrons per atom. In a cumulene, they occupy two degenerate $\pi$ bands, which results in a metal with two half-filled bands. However, such a system is unstable with respect to the Peierls transition \cite{Kertesz1978} in which unpaired electrons of neighbor carbon atoms form pairs resulting in the polyyne structure and a band gap is opened. It has been observed that behavior of finite carbon atomic strongly depends on the chain parity, i.e. whether the number of atoms in the chain is odd or even, \cite{Zanolli2010, Zhou2017, Lang1998, Cahangirov2010} which has been attributed to odd chains being closer to cumulenes and even chains to polyynes \cite{Zanolli2010}. The D1 and F polycyclic regions of 1D hybrid nanoobjects created by irradiation of GNRs are combined with even and odd chains, respectively \cite{Sinitsa2021}. It is, nevertheless, worth getting insight into the effect of chain parity keeping polycyclic regions the same. Therefore, we study here how the parity of carbon chains affects the structure and properties of the 1D hybrid nanoobjects with the F and D1 polycyclic regions.

\begin{figure}
	\centering
	\includegraphics[width=\columnwidth]{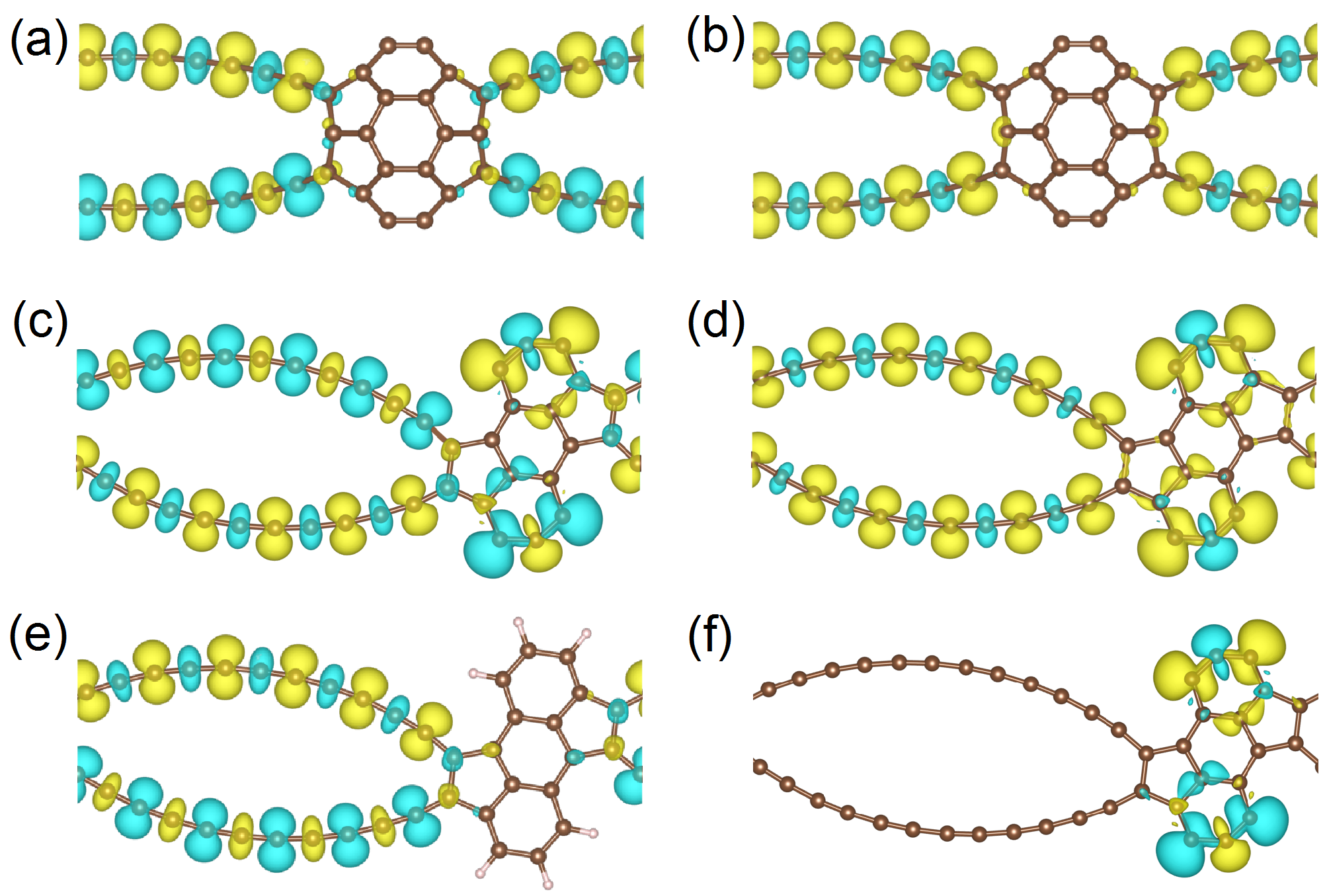}
	\caption{Spin maps computed using the PBE functional for 1D nanoobjects (isosurfaces 0.003 $e$/\AA$^3$) with antiparallel (AFM) and parallel (FM) spin ordering at opposite edges: (a) F-11 (AFM), (b) F-11 (FM), (c) D1-11 (AFM), (d) D1-11 (FM), (e) D1-11H (AFM) and (f) D1-12 (AFM). Carbon and hydrogen atoms are coloured in brown and white, respectively. One unit cell is shown.
	}
	\label{fig:spin}
\end{figure}

\begin{figure}
	\centering
	\includegraphics[width=\columnwidth]{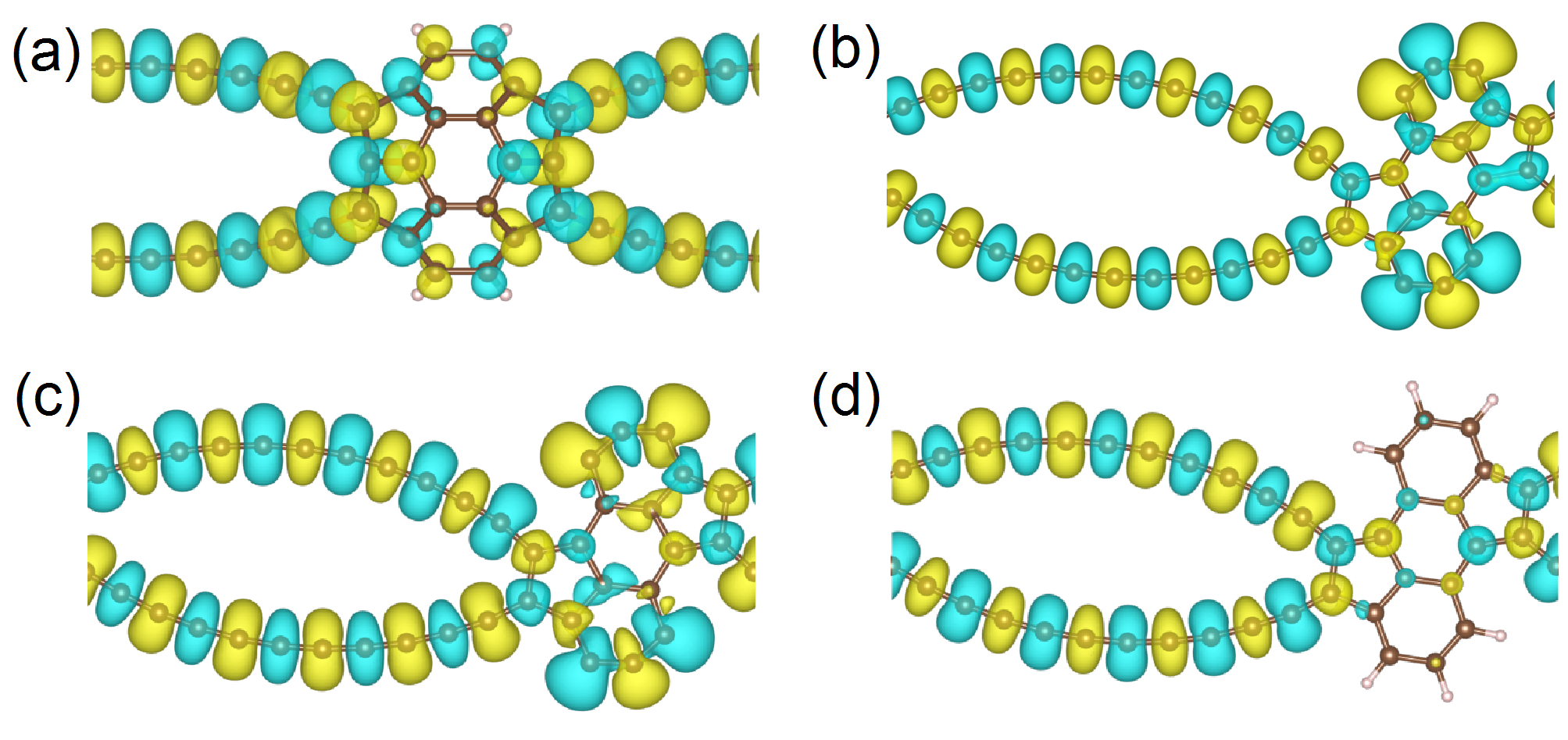}
	\caption{Spin maps computed using the HSE functional for 1D nanoobjects (isosurfaces 0.003 $e$/\AA$^3$) with antiparallel (AFM) and parallel (FM) spin ordering at opposite edges: (a) F-10H (FM), (b) D1-12 (AFM), (c) D1-11 (AFM) and (d) D1-11H (AFM). Carbon and hydrogen atoms are coloured in brown and white, respectively. One unit cell is shown.
	}
	\label{fig:spin_hse}
\end{figure}

First we describe the methods used for first-principles calculations. Then we present results on magnetism in the 1D nanoobjects, their structure and electronic properties. Finally conclusions are summarized and possible applications of nanoobjects are discussed.

\section{Methods}
Structural, magnetic and electronic properties of the 1D hybrid nanoobjects have been studied using the DFT calculations. It is known that semi-local exchange correlation functionals strongly underestimate band gaps \cite{Jain2011, Borlido2020, Heyd2004, Heyd2005}. Admixture of a portion of exact exchange and, correspondingly, account of non-local electron interactions in hybrid exchange-correlation functionals helps to alleviate this deficiency. Nevertheless, the calculations become much more heavy computationally. For this reason, we used the semi-local exchange and correlation functional of Perdew, Burke and Ernzerhof \cite{Perdew1996} (PBE) for geometry optimization of the nanoobjects and preliminary analysis of their magnetic and electronic properties. The latter were then refined using the screened exchange hybrid density functional of Heyd, Scuseria and Ernzerhof (HSE) \cite{Heyd2003, Heyd2006}, which reproduces experimental band gaps \cite{Heyd2004, Heyd2005} and optical excitation energies \cite{Barone2005a, Barone2005} for metallic and semiconducting single-walled carbon nanotubes. The comparison of the HSE and PBE results helps to get insight into the non-local effects arising from admixture of exact exchange. Note also that most of the computational data reported so far for GNRs and carbon chains were obtained using semi-local functionals and the data obtained using the PBE functional are useful to place the study in the context of previous works.

The DFT calculations were carried out using Quantum ESPRESSO \cite{Giannozzi2017, Giannozzi2009, QE}. The maximal kinetic energy of plane waves was 1088 eV. The kinetic energy cutoff for the charge density and potential was 4488 eV. The first-order Methfessel-Paxton smearing \cite{Methfesse1989} with the width of 0.07 eV was applied. The self-consistent field iterations were performed till the energy change in consecutive iterations became less than $10^{-9}$. For the PBE functional \cite{Perdew1996}, the projector augmented wave (PAW) method in combination with pseudopotentials from PS library \cite{DalCorso2014} was applied to the describe the interaction of valence electrons with the core. 6 k-points along the nanoobject axis were considered to converge the ground state density. The band structures were computed for 40 k-points along the nanoobject axis. Geometry optimization was performed till the residual forces of less than 0.05 eV/\AA, energy difference in successive optimization steps of 0.3 meV and stress of 10 MPa. For the HSE functional \cite{Heyd2003, Heyd2006}, the kinetic energy cutoff for the exact exchange operator was 1633 eV. 8 k-points along the nanoobject axis were considered. The norm-conserving pseudopotentials of Hartwigsen, Goedecker and Hutter \cite{Hartwigsen1998} were used. 

The ASAP-2024.0 (Atomistic Simulation Advanced Platform) package \cite{ASAP} was used to prepare the input files, launch the calculations and analyze the results. The figures of atomistic structures and spin maps were elaborated using VESTA \cite{Momma2011}.

\section{Results}
To investigate the effect of the polycyclic region structure on the nanoobject properties, we have performed the calculations for polycylic regions F and D1 (Fig. \ref{fig:geom}, see \cite{Sinitsa2021}) with and without hydrogen termination. The length of carbon chains was varied from 10 to 13 for the F region. For the D1 region, chains of length 11 and 12 were considered. The full list of nanoobjects studied is given in Table \ref{table:struct}. The nanoobjects with and without hydrogen are denoted as P-$m$H and P-$m$, respectively, where P is the polycyclic region (F or D1) and $m$ is the number of atoms in each chain.

Since the structure and electronic properties of the nanoobjects vary noticeably for different magnetic states, we start from the discussion of possible magnetic states and their stability. Then we consider structural and electronic properties for these states.

\begin{table*}
	\caption{Structure period $L$, bond length alternation $\delta$, band gap $\Delta$ and the k-point(s) corresponding to the band gap computed for 1D nanoobjects with parallel (FM), antiparallel (AFM) and no (NM) spin ordering at two edges using the PBE and HSE exchange-correlation functionals.}
	\centering
\renewcommand{\arraystretch}{0.95}
\begin{ruledtabular}
		\begin{tabular}{*{8}{c}}
			&  &  & &  \multicolumn{2}{c}{PBE} & \multicolumn{2}{c}{HSE} \\
			1D nanoobject & magnetic state & $L$ (\AA) & $\delta$ (pm) & k-point(s) & $\Delta$ (eV) & k-point(s) & $\Delta$ (eV) \\
			\hline
			F-10 & NM & 18.53 & 5.97 & $\Gamma$ & 0.22 & $\Gamma$ & 0.28 \\
			& FM\footnote{Magnetic states observed only for the HSE functional. The nanoobject geometry is assumed to be the same as for the NM state.} & -- & -- & -- & -- & $\Gamma$ & 1.19 \\
			F-10H & NM & 18.52 & 5.71 & $\Gamma$ & 0.34 & $\Gamma$ & 0.46 \\
			& FM$^\mathrm{a}$ & -- & -- & -- & -- & $\Gamma$ & 1.23 \\
			F-12H & NM & 20.80 & 5.46 & $X$ & 0.26 & $X$  & 0.33 \\
			& FM$^\mathrm{a}$ & -- & -- & -- & --  &  $X$  & 1.25 \\
			F-11 & NM & 19.75 & 3.09 & \multicolumn{2}{c}{metallic} & \multicolumn{2}{c}{metallic} \\
			& FM & 19.65 & 3.01 & \multicolumn{2}{c}{metallic} & $\Gamma$ & 0.19 \\
			& AFM & 19.64 & 2.99 & $\pi/(2L)$ & 0.11  & $\pi/(2L)$ & 0.76 \\
			F-11H & NM &  19.76 & 2.94 & \multicolumn{2}{c}{metallic} & \multicolumn{2}{c}{metallic}\\
			& FM &  19.66 & 2.87 & \multicolumn{2}{c}{metallic} & $\Gamma$ & 0.30 \\
			& AFM &  19.65 & 2.86 & $\pi/(2L)$ & 0.10 & $\pi/(2L)$ & 0.76 \\
			F-13H & NM & 22.23 & 2.91 & \multicolumn{2}{c}{metallic} & \multicolumn{2}{c}{metallic} \\
			& FM & 21.88 & 2.84 & \multicolumn{2}{c}{metallic} &  $X$ & 0.38 \\
			& AFM & 21.87 & 2.83 & $\pi/(2L)$ & 0.10 & $\pi/(2L)$ & 0.79 \\
			D1-11 & NM & 20.33 & 2.78 &  $X$ & 0.08 &  $X$ & 0.20  \\
			& FM & 20.36 & 3.02 &  $X$ & 0.21 &  $X$ & 0.22  \\
			& AFM & 20.33 & 2.78 &  $X$ & 0.36 &  $X$ & 1.00 \\
			D1-11H & NM & 20.35 & 2.56 &$\Gamma$ & 0.16 & $\Gamma$ & 0.22 \\
			& FM & 20.37 & 3.04 &  $X$ & 0.23 &  $X$ & 0.51 \\
			& AFM & 20.37 & 2.97 &  $X$ & 0.31 &  $X$ & 1.07 \\
			D1-12 & NM & 21.57 & 7.41 & \multicolumn{2}{c}{metallic} & \multicolumn{2}{c}{NA\footnote{Data not available because of convergence problems in exact exchange calculations related to numerical errors.}} \\
			& FM & 21.56 & 6.37 & \multicolumn{2}{c}{metallic} & $\Gamma$ --  $X$\footnote{Indirect band gap, k-points for the top of the valence band and bottom of the conduction band are indicated.} & 0.30 \\
			& AFM & 21.57 & 7.41 & $\Gamma$ --  $X^\mathrm{c}$ & 0.08 & $\Gamma$ --  $X^\mathrm{c}$ & 0.36 \\
			D1-12H & NM & 21.56 & 7.58 & $\Gamma$ --  $X^\mathrm{c}$  & 0.10 & $\Gamma$ --  $X^\mathrm{c}$  & 0.40 \\
		\end{tabular}
	\end{ruledtabular}	
	\label{table:struct}
\end{table*}

\subsection{Magnetism}
As known from previous studies, taking into account spin polarization is important for a proper description of GNRs and carbon chains. Spin-restricted calculations for zigzag GNRs give nearly flat bands at the Fermi level coming from edge states and featuring the electronic instability \cite{Son2006, Kunstmann2011, Magda2014, Cheng2012, Cheng2012prb, Song2010, Cantele2009, Pisani2007, Hod2007, Palacios2010}. The GNRs are stabilized by spin polarization at GNR edges. Spins are aligned in opposite directions for atoms of different sublattices and the magnitude of spin polarization is reduced fast away from the edge. Antiparallel spin ordering between two edges is the most energetically favourable for zigzag GNRs and results in opening a band gap. Parallel spin ordering at opposite edges is a bit higher in energy and leads to the metallic behavior but with no flat bands close to the Fermi level. Therefore, three magnetic states are normally considered for zigzag GNRs \cite{Son2006, Kunstmann2011, Magda2014, Cheng2012, Cheng2012prb, Song2010, Cantele2009, Pisani2007, Han2007}: non-magnetic (NM), antiferromagnetic (AFM) with aniparallel spin ordering between opposite edges and ferromagnetic (FM) with parallel spin ordering between opposite edges. It should be emphasized that both in the FM and AFM states, there is an alternation of spins for adjacent atoms, the terms FM and AFM refer here only to the relation between spin directions at opposite edges. Magnetism has been also discovered in odd carbon chains \cite{Zanolli2010, Cahangirov2010}. Alternating spin orientations are found for atoms along the chain.

Because of the above magnetic effects observed for GNRs and carbon chains, it is logical to expect similar phenomena for hybrid 1D nanoobjects. The relation between magnetism in the whole nanoobject and its components is the most evident when the semi-local PBE functional is used, while there are non-trivial effects for the hybrid HSE functional. Therefore, let us start from discussion of the PBE results. In our calculations, we set initial magnetic moments at edges of the polycylic region and in chains to induce convergence to the FM or AFM state. In spite of trying different initial guesses for the magnetic moments, we have not been able to get any noticeable magnetization in structures with the F region or hydrogen-terminated D1 region and even chains (Fig. \ref{fig:geom}) for the PBE functional. In structures with the same polycyclic regions but odd chains, only solutions with magnetization in chains but not at edges of the polycyclic region are obtained (Figs. \ref{fig:spin}a, b and e). In the structures with D1 region and even chains, the magnetization density is localized at edges of the polycyclic region (Fig. \ref{fig:spin}e). In the structures with D1 region without hydrogen termination and odd chains (Figs. \ref{fig:spin}c and d), spin ordering occurs both in chains and at the edges of the polycyclic region. 

Thus, according to the PBE functional, no spin ordering takes place at edges of the F polycyclic region (Figs. \ref{fig:spin}a and b). The edge carbon atoms of this region form a triple bond and no dangling bonds responsible for magnetism are left, the same as at armchair or reconstructed zigzag graphene edges \cite{Son2006, Gan2010, Koskinen2008, Kunstmann2011}. The D1 region, similar to graphene with pristine zigzag edges, has dangling bonds, which are stabilized by spin ordering \cite{Son2006, Kunstmann2011, Magda2014, Cheng2012, Cheng2012prb, Gan2010, Lee2009, Song2010} (Figs. \ref{fig:spin}c, d and f). Termination of the D1 region by hydrogen partially saturates the dangling bonds and the edge magnetization disappears (Fig. \ref{fig:spin}e). Note that in zigzag GNRs, partial or complete suppression of the edge magnetization is observed as well depending on hydrogen loading \cite{Kunstmann2011, Gan2010, Wassmann2008}.

As for magnetism in carbon chains, the PBE functional gives only small magnetic moments at atoms of even chains (Fig. \ref{fig:spin}e, see also \cite{Sinitsa2021}), while in odd chains, magnetic moments comparable to those at the edges of D1 polycyclic region are observed (Figs. \ref{fig:spin}a--e). Such a dependence on the chain parity is in agreement with calculations for carbon chains attached to graphene ribbons \cite{Zanolli2010} and isolated carbon chains \cite{Cahangirov2010} and is related to the occupation of bands of the chains. In even chains, the bands are fully filled, all bonds are saturated and consequently no magnetization is observed. In odd chains, there are half-filled bands leading to non-negligible magnetic moments. 

As seen from the above, in the calculations using the PBE functional, the magnetic properties of 1D hybrid nanoobjects are determined in the trivial way by those of their components. This is not the case for the hybrid HSE functional. The latter takes into account non-local interactions through admixture of exact exchange and this allows to reveal synergistic effects of the nanoobject components. The calculations using the hybrid HSE functional show magnetism in the nanoobjects with F region and even chains (Fig. \ref{fig:spin_hse}a), which are non-magnetic according to the PBE functional. This magnetism is stable with respect to hydrogen termination of the polycyclic region and significant magnetic moments are observed both at atoms of the chains and in the F region. Although the total magnetic moment for this state is zero, we denote it as FM because the magnetic moments of atoms are aligned in the same direction at opposite nanoobject edges. Considerable magnetic moments are also obtained using the HSE functional for atoms in even chains attached to the D1 region without hydrogen termination (Fig. \ref{fig:spin_hse}b), different from the PBE result (Fig. \ref{fig:spin}f). Note, however, that the HSE functional still gives negligible magnetic moments for atoms in the nanoobject with the hydrogen terminated D1 region and even chains. Magnetization is also low in the F region connected to odd chains, the same as for the PBE functional (Figs. \ref{fig:spin}a and b). Therefore, new magnetic effects coming from the account of the exact exchange contribution are manifested only for specific combinations of chains and polycyclic regions. 

The distributions of the magnetization density in nanoobjects with odd chains obtained using the HSE functional are qualitatively similar to those for PBE (compare Figs. \ref{fig:spin_hse}c and d and Figs. \ref{fig:spin}c and e, respectively), i.e. magnetism in these structures is described by the PBE functional more or less well. It should be noted, however, that magnitudes of magnetic moments for atoms in odd chains do not vary that much for adjacent atoms according to the HSE functional as compared to PBE. 

The energy difference between the FM and AFM states and the relative energy of the NM state with respect to the ground state computed for the nanoobjects using the PBE and HSE functionals are listed in Table \ref{table:edif}. These energies characterize the strength of magnetic instability and magnetic interaction, respectively \cite{Pisani2007, Cantele2009}. For all the considered nanoobjects with odd chains or polycyclic regions with dangling bonds, AFM and FM states are much more stable than the NM state with the magnetic interaction energy in the range of 0.1 -- 0.3 eV per unit cell for the PBE functional and 0.7 -- 1.8 eV per unit cell for the HSE functional (Table \ref{table:edif}). In most of the cases, the AFM state is the most energetically favorable according both to the PBE and HSE functionals (Table \ref{table:edif}). For the D1-12 nanoobject, the PBE functional gives the FM ground state, which is more stable than the AFM state by 17 meV per unit cell. Qualitatively similar results were obtained in our previous calculations for the D1-16 nanoobject \cite{Sinitsa2021}. However, the account of non-local effects in the HSE functional stabilizes the AFM state and makes it slightly more preferable energetically than the FM one (Table \ref{table:edif}). The relative energies of the NM states for the nanoobjects with the F region and even chains in which the magnetism is induced exclusively by non-local effects described by the HSE functional are much smaller, 0.1 -- 0.2 eV per unit cell. No AFM states have been found for these structures. 

Experimental measurements \cite{Magda2014} for doped zigzag GNRs at room temperature showed that the GNRs of width that is less than 7 nm behave as semiconductors, while those wider than 7 nm as metals. This change in the GNR behavior upon increasing the width was attributed to the transition from the AFM to FM state. Therefore, this study provides an evidence that the magnetic order can be observed in zigzag GNRs at room temperature. Note that calculations with the PBE functional suggest that the energy difference between the FM and AFM states, i.e. the magnetic interaction strength, in such ribbons can be on the order of 1 meV per unit cell \cite{Son2006, Cantele2009, Pisani2007} and the energy penalty for the NM state, i.e. the magnetic instability strength, is about 80 meV per unit cell \cite{Cantele2009}. For the hybrid B3LYP functional, the corresponding values are $\sim$3 meV and 300 meV per unit cell, respectively \cite{Pisani2007}. This is considerably smaller than similar energies for the nanoobjects considered here and gives hope that magnetic effects can be observed for them as well.

The magnetic moments per unit cell computed for hybrid nanoobjects in the FM state are given in Table \ref{table:edif} in units of Bohr magneton, $\mu_\mathrm{B}$. The magnetic moment per unit cell is about $2\mu_\mathrm{B}$ for nanoobjects with the F polycyclic region or hydrogenated D1 polycyclic region without dangling bonds and odd chains independent of the chain length (Fig. \ref{fig:spin}b). Thus, the magnetic moment per an odd chain is $\mu_\mathrm{B}$. In the case of the D1 polycyclic region without hydrogen (Fig. \ref{fig:spin}d), the magnetic moment per unit cell is increased additionally by $2\mu_\mathrm{B}$. In the case of D1-12 nanoobject, the magnetization density is localized on the edges of the polycyclic region (similar to Fig. \ref{fig:spin}f) and corresponds to $3\mu_\mathrm{B}$ according to the PBE functional. In the case of HSE functional, magnetization is additionally induced in the chains (similar to Fig. \ref{fig:spin_hse}b) and the resulting magnetic moment becomes $2\mu_\mathrm{B}$ per unit cell. The total magnetic moments of the FM states of the nanoobjects with even chains and F polycyclic regions (Fig. \ref{fig:spin_hse}a) are exactly zero. The magnetization density for these nanoobjects is symmetric up to the sign change with respect to the mirror plane passing through the middle of the F region perpendicular to the plane of Fig. \ref{fig:spin_hse}a and nanoobject axis. 

\begin{table*}
	\caption{Calculated energy differences $\Delta E_\mathrm{FM}$ (in eV per unit cell) between states of 1D hybrid nanoobjects with parallel (FM) and antiparallel (AFM) spin ordering at two edges, relative energies $\Delta E_\mathrm{NM}$ (in eV per unit cell) of states with no spin ordering (NM) with respect to the ground states and magnetic moments $\mu$ (in $\mu_\mathrm{B}$ per unit cell) in the FM states.}
	\centering
	\renewcommand{\arraystretch}{0.8}
	\begin{ruledtabular}
		\begin{tabular}{*{7}{c}}
			&  \multicolumn{3}{c}{PBE} & \multicolumn{3}{c}{HSE} \\ 
			1D nanoobject &  $\Delta E_\mathrm{FM}$ (eV) & $\Delta E_\mathrm{NM}$ (eV) & $\mu$ ($\mu_\mathrm{B}$)  &  $\Delta E_\mathrm{FM}$ (eV) & $\Delta E_\mathrm{NM}$ (eV) & $\mu$ ($\mu_\mathrm{B}$) \\\hline
			F-10\footnote{No magnetic states observed for the PBE functional. No AFM state found for the HSE functional.} & -- & -- & -- & -- & 0.155 & 0.00 \\ 
			F-10H$^\mathrm{a}$ &   -- & -- & -- & -- & 0.135 & 0.00 \\
			F-12H$^\mathrm{a}$ &  -- & -- & -- & --  & 0.199 & 0.00 \\ 
			F-11 & 0.021 & 0.119 & 2.03 & 0.108 & 0.843 & 2.00 \\
			F-11H & 0.016 & 0.153 & 2.04 & 0.066 & 0.887 & 2.00 \\
			F-13H & 0.016 & 0.123 & 2.03 & 0.046 & 0.889 & 2.00 \\
			D1-11 & 0.125 & 0.265 & 4.03 & 0.262 & 1.809 & 3.99 \\
			D1-11H & 0.018 & 0.174 & 2.04 & 0.201 & 1.004 & 2.00 \\
			D1-12 & -0.017 & 0.289 & 3.00 & 0.003 & NA\footnote{Data not available because of convergence problems in exact exchange calculations
				related to numerical errors.} & 2.00\\
		\end{tabular}
	\end{ruledtabular}
	\label{table:edif}
\end{table*}

\subsection{Structure}
Examples of optimized structures of the hybrid nanoobjects are shown in Figs. \ref{fig:geom} and \ref{fig:spin}. The periods $L$ of optimized nanoobjects are listed in Table \ref{table:struct}. The changes in the periods for different magnetic states are within 10 pm. The differences in the bond lengths for the AFM and FM states are normally within 1 pm. Note, however, that sometimes there are significant qualitative changes in the structure of hybrid nanoobjects for different magnetic states. For example, spin-restricted geometry optimization of the nanoobjects with the F polycyclic region and odd chains leads to a structure geometry with perfectly parallel chains, while account of spin polarization results in spatial separation of the chains as shown in Fig. \ref{fig:spin}a and b. 

For nanoobjects with the polycyclic region F (Figs. \ref{fig:geom}a--c and \ref{fig:spin}a and b), the optimized structures are symmetric with respect to two mirror planes perpendicular to the plane of Figs. \ref{fig:geom}a--c and \ref{fig:spin}a and b and crossing in the center of the polycyclic region: one aligned along the nanoobject axis and the other perpendicular to it. The differences in lengths of bonds in carbon chains equivalent with respect to these two mirror planes are negligibly small (do not exceed 0.03 pm). The nanostructures are not flat. Since the polycyclic region F consists of pentagons and hexagons, it has a cap shape similar to parts of fullerenes or nanotube tips (Fig. \ref{fig:geom}b).

The structures with the polycyclic region D1 (Figs. \ref{fig:geom}d and \ref{fig:spin}c--f) are flat. These nanoobjects are symmetric with respect to rotation around $C_2$ axis perpendicular to the plane of Figs. \ref{fig:geom}d and \ref{fig:spin}c--f and passing through the center of the polyciclic region. The differences in lengths of bonds of two chains equivalent with respect to this rotation are negligible. At the same time, non-negligible differences on the order of 1 pm are observed in lengths of bonds of the same chain at the same distance from the middle points of the chains.

Hydrogen termination of edges of polycyclic regions (Figs. \ref{fig:geom}c and d and \ref{fig:spin}e) leads to changes in bond lengths only close to the edges. In the nanoobjects with the polycyclic region F without hydrogen, the edge carbon atoms form a short triple bond of length 1.27 \AA. When hydrogen is added, this bond is increased in length to 1.43 -- 1.44 \AA. In the structures with the polycyclic region D1, the length of bonds between edge atoms is 1.28 -- 1.31 \AA{} and they are increased by 8 -- 12 pm upon hydrogen addition. The bond lengths in chains are modified at most by 0.3 pm when hydrogen is attached.

Both odd and even carbon chains within 1D nanoobjects show a bond length alternation along the chain. To characterize it, we have computed the amplitude of bond variation according to   \cite{Ravagnan2009, Zanolli2010} as
\begin{equation} \label{bond_alternation}
\delta=\frac{1}{2}\left\vert{\frac{1}{n_e}\sum_{j=1}^{n_e}\left(d_{2j-1}+d_{n-(2j-1)}\right) - \frac{1}{n_0}\sum_{j=1}^{n_0}\left(d_{2j}+d_{n-2j}\right)}\right\vert,
\end{equation}
\noindent where $d_i=|r_i - r_{i+1}|$, $n$ is the number of atoms in the chain, $n_e$ is the integer part of $(n+2)/4$ and $n_0$ is the integer part of $n/4$. The end bonds of the chains are not included. The computed values of $\delta$ are $\sim$3 pm for odd chains and $\sim$6 -- 8 pm for even chains (Table \ref{table:struct}). This smaller bond length alternation in odd chains indicates that they are closer to cumulene structure, different from even chains that have a more pronounced polyyne character. Indeed in even chains there is a perfect alternation between single and triple bonds. In odd chains, the character of the bonds changes from single-triple pairs at the chain ends to equal double bonds in the middle imposed by the symmetry considerations. Note that similar values of bond length alternation of 5 -- 9 pm and 3 -- 3.5 pm were obtained for chains of 8 and 9 atoms, respectively,  attached to graphene fragments in \cite{Zanolli2010, Ravagnan2009}. The values of 6 -- 8 pm were found for 8-atom chains in carbon nanobracelets \cite{Vyrko2024}. Our previous calculations \cite{Sinitsa2021} for the D1-16 nanoobject gave the bond length alternation of about 6 pm. Values of similar magnitude were obtained for bond length alternation in carbon nanotubes \cite{Poklonski2012}.

\subsection{Electronic properties}
\begin{figure*}
	\centering
	\includegraphics[width=0.95\textwidth]{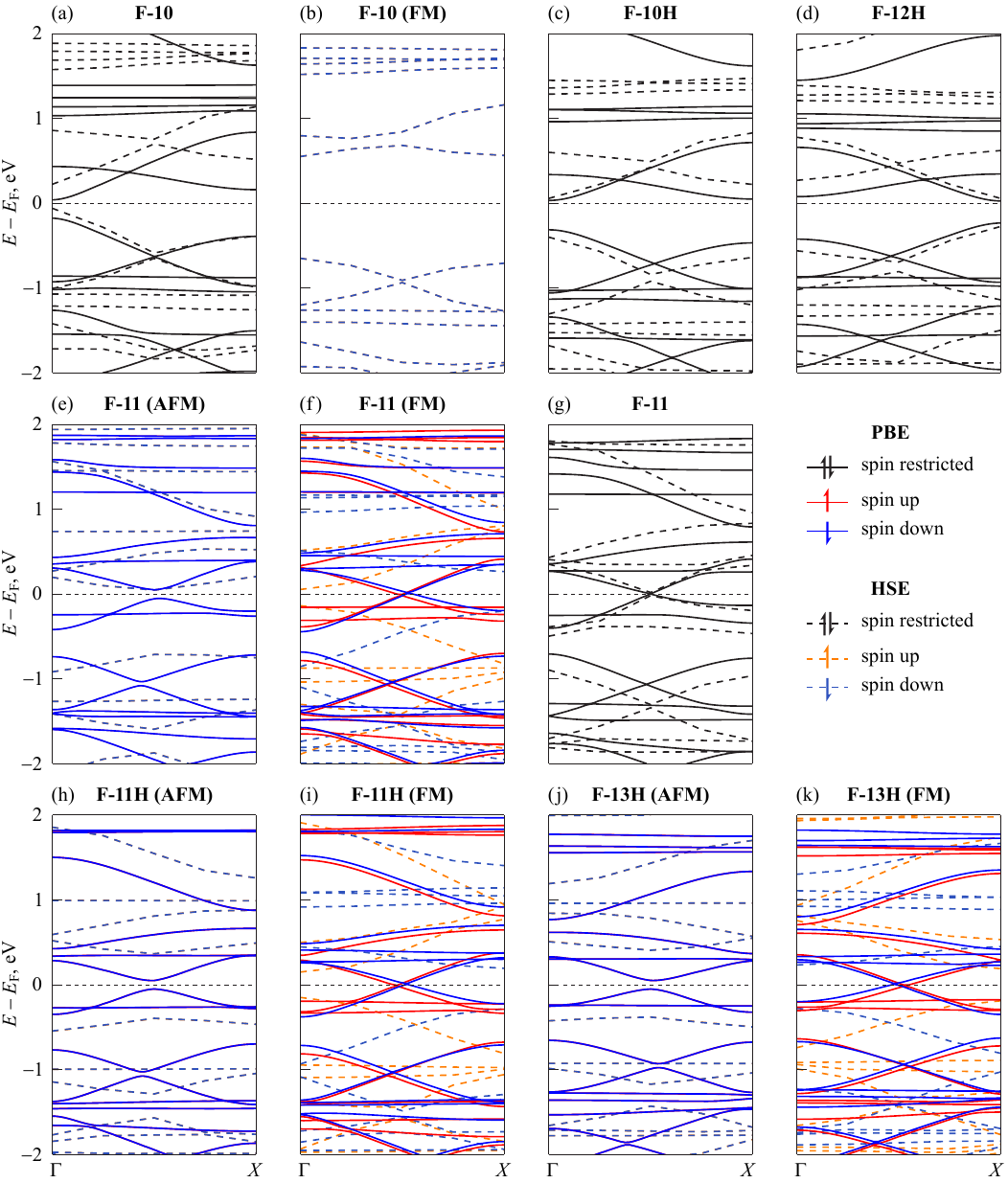}
	\caption{Computed band structures of 1D hybrid nanoobjects with the F polycyclic region:  (a) F-10 (NM), (b) F-10 (FM), (c) F-10H (NM), (d) F-12H (NM), (e) F-11 (AFM), (f) F-11 (FM), (g) F-11 (NM), (h) F-11H (AFM), (i) F-11H (FM), (j) F-13H (AFM), (k) F-13H (FM). The red/orange and blue/dark blue lines correspond to spin up and down, respectively. The black lines are obtained by spin-restricted calculations. The results of the calculations using the PBE and HSE functionals are shown by solid and dashed lines, respectively.
	}
	\label{fig:bands_F}
\end{figure*}

Band structures calculated for the nanoobjects with the polycyclic region F are shown in Fig. \ref{fig:bands_F}. According to the calculations using the PBE functional, the nanoobjects with the F region and even chains (Figs. \ref{fig:bands_F}a, c and d) are semiconductors with a direct band gap of 0.2 -- 0.3 eV at $\Gamma$ or $X$ point (which correspond to coordinates 0 and $\pi/L$ along the nanoobject axis in the reciprocal space, respectively, with $L$ being the nanoobject period). The computed band gaps are listed in Table \ref{table:struct}. Admixture of exact exchange in the HSE functional leads for such nanoobjects only to a small band gap increase in spin-restricted calculations. However, in spin-polarized calculations, it also induces magnetization both in the even chains and polycyclic region. As a result, the band gap for the ground state of this nanoobject  becomes about 1.2 eV (Fig. \ref{fig:bands_F}b, Table \ref{table:struct}).

Spin-restricted calculations for the nanoobjects with the polycyclic region F and odd chains give a metallic band structure (Fig. \ref{fig:bands_F}g). This is again related to the fact that all bonds are saturated in even chains, while there are half-filled bands in odd chains. Note that this behavior is independent of the exchange-correlation functional used. Although the nanoobjects with odd chains are unstable with respect to spin ordering at atoms of the chains, in the calculations with the PBE functional, the band structure stays rather close to the spin-restricted case (Figs. \ref{fig:bands_F}e and f). In the FM state, the nanoobjects are metallic (Figs. \ref{fig:bands_F}f, i and k). There is a difference in energies of localized states for spin up and down but dispersive bands are very close. In the most energetically favourable AFM state (Figs. \ref{fig:bands_F}e, h and j), the bands for spin up and down are the same and a small gap of about 0.10 eV is opened (Table \ref{table:struct}). In the calculations with the HSE functional, the band gap for the AFM state increases by 0.7 eV (Figs. \ref{fig:bands_F}e, h and j). The band structure for the FM state changes even more dramatically and a gap of about 0.2 eV is opened at $\Gamma$ point (Figs. \ref{fig:bands_F}f, i and k). Still this gap is significantly smaller than the one for the AFM state (Table \ref{table:struct}). 

\begin{figure*}
	\centering
	\includegraphics[width=0.95\textwidth]{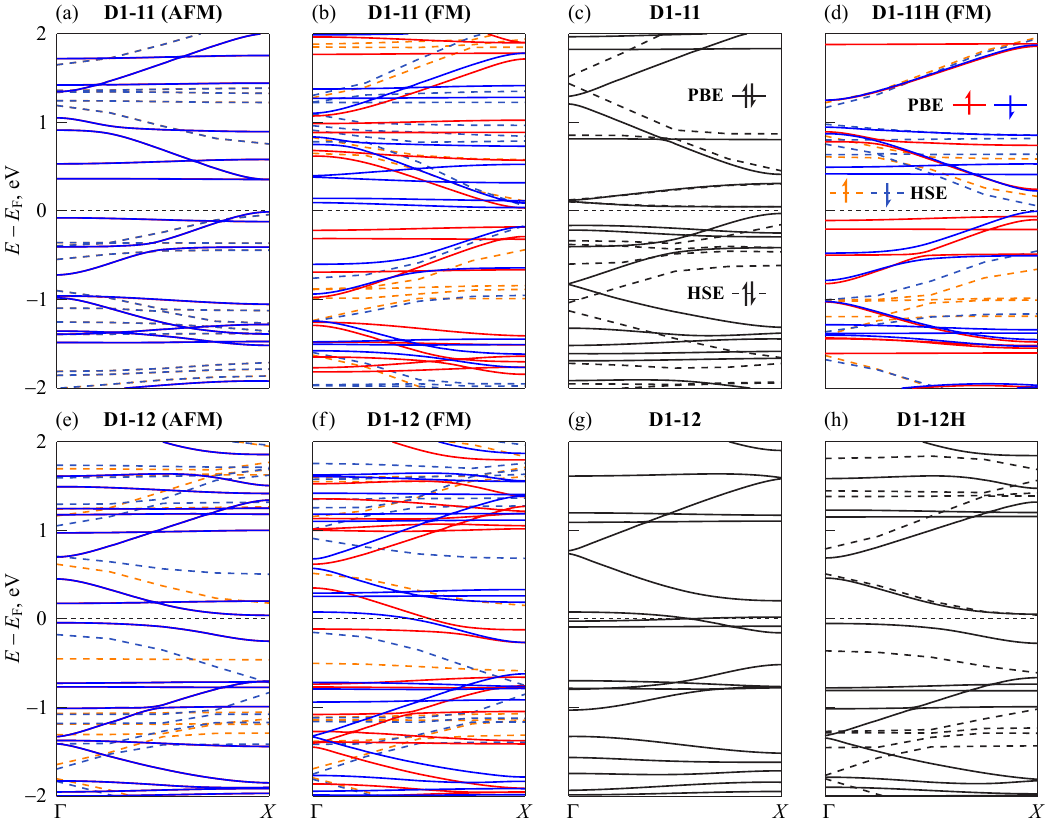}
	\caption{Computed band structures of 1D hybrid nanoobjects with the D1 polycyclic region: (a) D1-11 (AFM), (b) D1-11 (FM), (c) D1-11 (NM), (d) D1-11H (FM), (e) D1-12 (AFM), (f) D1-12 (FM), (g) D1-12 (NM), (h) D1-12H (NM). The red/orange and blue/dark blue lines correspond to spin up and down, respectively. The black lines are obtained by spin-restricted calculations. The results of the calculations using the PBE and HSE functionals are shown by solid and dashed lines, respectively.
	}
	\label{fig:bands_D1}
\end{figure*}

Hydrogen termination of the polycyclic regions leads only to minor changes in the band structure for the nanoobjects with the F region (Figs. \ref{fig:bands_F}c, h and i). For the F-10 nanoobject (Figs. \ref{fig:bands_F}a and c), the band gap is increased at $\Gamma$ point and slightly decreased at $X$ point (Table \ref{table:struct}). The band structures for the F-11H and F13-H nanoobjects (Figs. \ref{fig:bands_F}h--k) look similar to that for the F-11 nanoobject (Figs. \ref{fig:bands_F}e and f), especially close to the Fermi level (see also Table \ref{table:struct}).

Band structures of the nanoobjects with the polycyclic region D1 are shown in Fig. \ref{fig:bands_D1}. Unpaired electrons from edges of the D1 region fill the initially half-filled bands of odd chains and the corresponding nanoobject D1-11 results to be semiconducting (Figs. \ref{fig:bands_D1}a--d). The nanoobject has a direct band gap at $X$ point. It is rather small in spin-restricted calculations but increases when spin ordering is taken into account (Table \ref{table:struct}). Admixture of exact exchange in the HSE functional leads to the increase of the band gap by 0.64 eV in the AFM state (Fig. \ref{fig:bands_D1}a). Changes in the band gap for the FM and NM states upon including exact exchange are not that significant but even for these states, flat bands are moved farther away from the Fermi level (Figs. \ref{fig:bands_D1}b--d). Again the main difference in band structures for spin up and down in the FM state and the effect of hydrogen termination are related to positions of flat bands (Figs. \ref{fig:bands_D1}b and d).

The spin-restricted calculations for the D1-12 nanoobject give flat bands at the Fermi energy that correspond to dangling bonds at edges of the D1 region (Fig. \ref{fig:bands_D1}g). Spin ordering stabilizes the system and the corresponding bands move away from the Fermi level (Figs. \ref{fig:bands_D1}e and f). According to the calculations with the PBE functional, the structure becomes semiconducting in the AFM state with a small indirect band gap of 0.08 eV between $\Gamma$ and $X$ points (Fig. \ref{fig:bands_D1}e, Table \ref{table:struct}). In the FM state (Fig. \ref{fig:bands_D1}f), there are conduction bands crossing the Fermi level. These results fully agree with the previous calculations for the D1-16 nanoobject \cite{Sinitsa2021} giving the indirect band gap of 0.06 eV for the AFM state and metallic FM state. In the calculations for the D1-12 nanoobject with the HSE functional, the indirect band gap in the AFM state becomes 0.38 eV (Fig. \ref{fig:bands_D1}e) and the band gap of 0.30 eV is opened in the FM state (Fig. \ref{fig:bands_D1}f). Note that non-local effects taken into account through the HSE functional lead to splitting of bands for spin up and down not only in the FM but also in the AFM state. Actually the AFM and FM states in this case are very close in energy (Table \ref{table:edif}) and have similar band structures (Figs. \ref{fig:bands_D1}e and f). In both cases shifting the Fermi level by 0.2 eV can drive the system to the state that behaves as a metal for one spin orientation and a semiconductor for the other. This can be achieved by chemical doping or using a gate. The termination of the D1 region by hydrogen saturates the dangling bonds and the nanoobject becomes a semiconductor with the indirect band gap of 0.10 eV and 0.40 eV according to the PBE and HSE functionals, respectively (Fig. \ref{fig:bands_D1}h).

\begin{figure*}
	\centering
	\includegraphics[width=\textwidth]{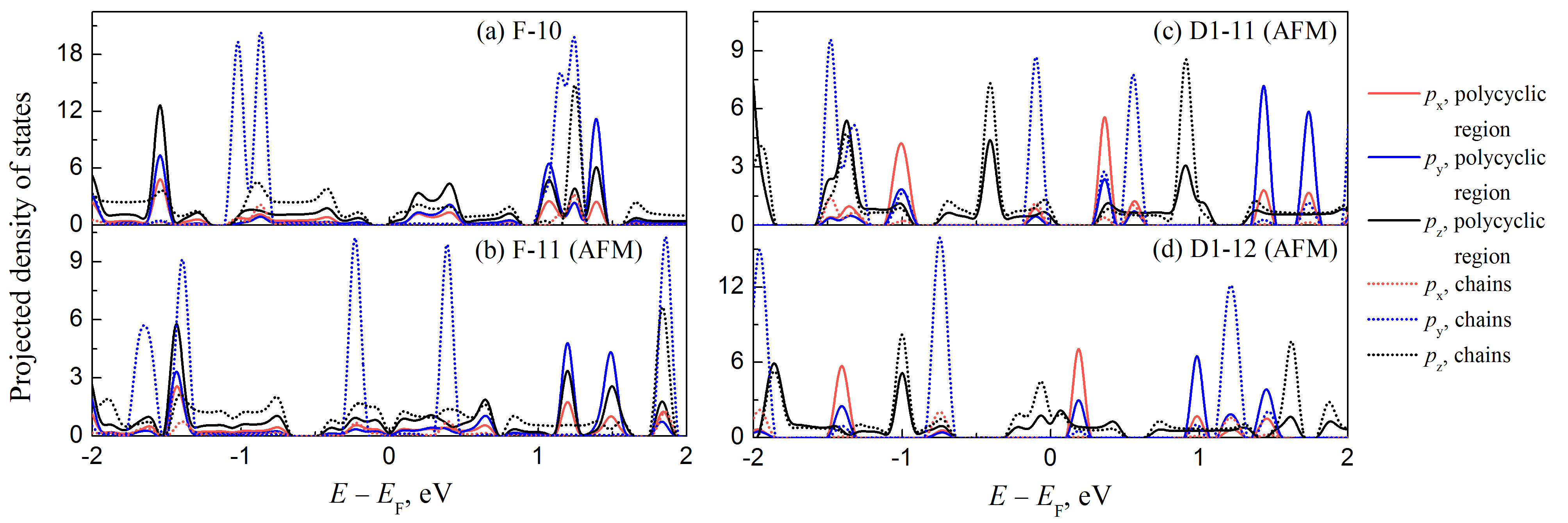}
	\caption{Projected density of states (in arbitrary units) obtained using the PBE functional as a function of energy $E-E_\mathrm{F}$ (in eV) relative to the Fermi level for (a) F-10 (NM), (b) F-11 (AFM), (c) D1-11 (AFM) and (d) D1-12 (AFM) nanoobjects. The projections on the $p_\mathrm{x}$, $p_\mathrm{y}$ and $p_\mathrm{z}$ orbitals are shown in red, blue and black, respectively ($x$, $y$ and $z$ axes are directed along the nanoobject axis and perpendicular in-plane and out-of-plane directions for Figs. \ref{fig:geom} and \ref{fig:spin}). Solid and dotted lines correspond to the projections on atoms of the polycylic region and chains, respectively.
	}
	\label{fig:pdos}
\end{figure*}

The analysis of projected density of states (PDOS) (Fig. \ref{fig:pdos}) reveals that the major contribution to the bands with a significant dispersion, including those close to the Fermi level, corresponds to $p_z$ orbitals of atoms in the chains and in the polycyclic region ($x$, $y$ and $z$ axes are directed along the nanoobject axis and perpendicular in-plane and out-of-plane directions for Figs. \ref{fig:geom} and \ref{fig:spin}). While in flat structures with the D1 polycyclic region, other contributions to the dispersive bands are negligible (Figs. \ref{fig:pdos}c and d), in nanoobjects with the F region, which are not flat, there are also contributions from $p_x$ and $p_y$ orbitals (Figs. \ref{fig:pdos}a and b). The flat bands composed of $p_y$ and $p_z$ orbitals of the chains, $p_x$, $p_y$ and $p_z$ orbitals of the polycyclic region are observed in the energy interval from -2 eV to 2 eV from the Fermi level.

\section{Conclusions}
We have performed spin-polarized DFT calculations to investigate the properties of 1D hybrid nanoobjects consisting of polycyclic hydrocarbon regions alternating with double carbon chains. The calculations show that structural, magnetic and electronic properties of such nanoobjects can be changed drastically by modifying the chain parity and edge structure of polycyclic regions. Depending on these parameters, the nanoobjects can be promising for different spintronic applications. We have found a significant splitting of bands for spin up and down in the ground state of the nanoobjects with the D1 polycyclic region having dangling bonds and even chains, like D1-12 or D1-16 (see \cite{Sinitsa2021}). Shifting the Fermi level by chemical doping or using a gate can make these nanoobjects metallic for one spin component and semiconducting for the other. These structures thus belong to magnetic semiconductors, materials that have been searched for already a long time \cite{Methfessel1968, Pappas2013, Ohno1996, Kalita2023, Cheng2022}. The high interest to magnetic semiconductors is explained by their potential to generate spin-polarized currents and easiness of integration into semiconductor devices \cite{Wolf2001, Kioseoglou2004, Telegin2022, Zutic2004}. For the nanoobjects with odd chains, we observe large differences in the band gaps for the magnetic states with parallel and antiparallel spin ordering at opposite edges. Therefore, such nanoobjects can be used in magnetic tunnel junctions, which find application in magnetic field sensors, read heads for hard drives, galvanic isolators, and magnetoresistive random access memory \cite{Wolf2001, Zutic2004, Peng2014, Zhang2021a}.

Our calculations reveal highly stable magnetic states for 1D nanoobjects corresponding to spin ordering at edges of polycyclic regions having dangling bonds and in odd chains. When non-local electron interactions are taken into account by using the hybrid HSE exchange-correlation function, magnetization is also induced in even chains and polycyclic region F. Note that the latter phenomenon depends on the combination of specific chain parity and polycyclic region structure, i.e. it is a result of synergy of nanoobject components. For the nanoobjects including odd chains and/or polycyclic regions with dangling bonds, the AFM state with antiparallel spin ordering in two chains and edges of the polycyclic region corresponds to the ground state. However, very close energies are obtained for the FM state with parallel spin ordering and AFM state for the nanoobjects consisting of the polycyclic regions having dangling bonds and even chains (like D1-12). The ground state of the nanoobjects with the F polycyclic region and even chains has parallel spin ordering at opposite edges (according to the HSE functional) but the total magnetic moment is zero because of symmetry considerations.

There is an oscillation in the electronic properties of the nanoobjects upon increasing the chain length, while the nanoobjects with chains of the same parity but different length behave similarly. Modification of the edge structure of polycyclic regions can lead to minima-maxima switching in the band gap dependence on the chain length and changes in the band gap character (e.g., from direct to indirect). If the polycylic region has no dangling bonds like the region F, non-magnetic structures with even chains are semiconducting with a moderate gap (0.1 -- 0.4 eV). This gap becomes as large as 1.2 eV for magnetic states observed using the hybrid exchange-correlation functional. At the same time, the structures with odd chains and polycyclic regions without dangling bonds are metallic or have only a small gap in the NM and FM states, while the ground AFM state is semiconducting (with the gaps of 0.1 and 0.8 eV for the PBE and HSE functionals, respectively). The opposite occurs in the structures in which the polycylic region has dangling bonds (like D1). The nanoobjects with even chains are metallic or have a small gap in the NM and FM states, while the AFM state has a moderate gap. The nanoobjects with the polycylic region having dangling bonds and odd chains have a moderate band gap already in the NM and FM states, which is further increased in the ground AFM state. 

Hydrogen termination of edges of the polycyclic region usually suppresses the magnetization in the polycyclic region. Structural changes upon hydrogen adsorption are limited to edges of the polycyclic region. The effect of hydrogen termination on the band structure is mostly reduced to moving flat bands farther away from the Fermi level. 

The amplitude of bond length alternation is found to be about twice smaller for odd chains as compared to even ones. This is explained by the fact that single-triple bond alternation is frustrated in the middle of odd chains as imposed by symmetry \cite{Ravagnan2009, Zanolli2010, Cahangirov2010}. In this sense, odd chains have a less pronounced polyyne character and, as a consequence, atoms of the odd chains have non-negligible magnetic moments independent of the structure of the polycyclic region \cite{Zanolli2010}. Nevertheless, the nanoobjects both with odd and even chains can be metallic or semiconducting depending on the structure of the polycyclic region. This is a demonstration of the fact that the chain structure alone does not determine the electronic properties and it is needed to consider occupations of bands of the nanoobject as a whole \cite{Lang1998}.

As discussed above, taking into account non-local electron interactions by using the hybrid HSE functional leads to important corrections to the results on magnetic and electronic properties of 1D nanoobjects. The band gap of the nanoobjects is opened for metallic nanoobjects or increased by up to 1 eV for semiconducting ones, while the magnetic states are stabilized by 0.7 -- 1.5 eV. Additionally, the hybrid functional predicts magnetism in nanoobjects that are non-magnetic according to the semi-local functional like those with even chains and polycyclic region F or region D1 without hydrogen termination. Therefore, account of non-local effects, e.g., via the use of the hybrid exchange-correlation functional, is necessary for adequate description of nanoobject properties and synergistic effects of nanoobject components. 

We expect that our results showing promise of 1D nanoobjects for spintronic applications as well as recent advances in atomically-precise synthesis of GNRs \cite{Cai2010, Zhou2020, Brede2023}, which can serve as precursors for nanoobject production \cite{Sinitsa2021}, will stimulate experimental research on such hybrid systems.


\subsection*{Acknowledgements}
	These studies were performed using ASAP-2024.0 — Atomistic Simulation Advanced Platform by Simune Atomistics S.L. I.V.L acknowledges the IKUR HPC project "First-principles simulations of complex condensed matter in exascale computers" funded by MCIN and by the European Union NextGenerationEU/PRTR-C17.I1, as well as by the Department of Education of the Basque Government through the collaboration agreement with nanoGUNE within the framework of the IKUR Strategy, computer resources at MareNostrum and the technical support provided by Barcelona Supercomputing Center (RES grant nos. FI-2022-1-0023, FI-2022-2-0035, FI-2022-3-0048 and FI-2023-1-0037). A.S.S. and A.A.K. acknowledge computing resources of the federal
	collective usage center Complex for Simulation and Data Processing
	for Megascience Facilities at NRC ``Kurchatov Institute", https://ckp.nrcki.ru/. A.M.P., and Y.E.L. acknowledge the support by the Russian Science Foundation grant No. 23-42-10010, {https://rscf.ru/en/project/23-42-10010/}. S.A.V. and N.A.P. acknowledge support by the Belarusian Republican Foundation for Fundamental Research (Grant No. F23RNF-049) and by the Belarusian National Research Program ``Convergence-2025". 

\section*{Data Availability Statement}
The raw data required to reproduce these findings are available to
download from \url{https://doi.org/10.5281/zenodo.12184530}.

\bibliography{refs}

\begin{thebibliography}{100}
\expandafter\ifx\csname url\endcsname\relax
  \def\url#1{\texttt{#1}}\fi
\expandafter\ifx\csname urlprefix\endcsname\relax\def\urlprefix{URL }\fi
\expandafter\ifx\csname href\endcsname\relax
  \def\href#1#2{#2} \def\path#1{#1}\fi

\bibitem{Geim2009}
A.~K. Geim, K.~S. Novoselov, The rise of graphene, Word Scientific: Singapore,
  2009, pp. 11--19.
\newblock \href {https://doi.org/https://doi.org/10.1142/9789814287005_0002}
  {\path{doi:https://doi.org/10.1142/9789814287005_0002}}.

\bibitem{Geim2009a}
A.~K. Geim, Graphene: Status and prospects, Science 324 (2009) 1530--1534.
\newblock \href {https://doi.org/https://doi.org/10.1126/science.1158877}
  {\path{doi:https://doi.org/10.1126/science.1158877}}.

\bibitem{Sato2015}
S.~Sato, Graphene for nanoelectronics, Japanese Journal of Applied Physics 54
  (2015) 040102.
\newblock \href {https://doi.org/https://doi.org/10.7567/JJAP.54.040102}
  {\path{doi:https://doi.org/10.7567/JJAP.54.040102}}.

\bibitem{Westervelt2008}
R.~M. Westervelt, Graphene nanoelectronics, Science 320 (2008) 324--325.
\newblock \href {https://doi.org/https://doi.org/10.1126/science.1156936}
  {\path{doi:https://doi.org/10.1126/science.1156936}}.

\bibitem{Banhart2011}
F.~Banhart, J.~Kotakoski, A.~V. Krasheninnikov, Structural defects in graphene,
  ACS Nano 5 (2011) 26--41.
\newblock \href {https://doi.org/10.1021/nn102598m}
  {\path{doi:10.1021/nn102598m}}.

\bibitem{Terrones2012}
H.~Terrones, R.~Lv, M.~Terrones, M.~S. Dresselhaus, The role of defects and
  doping in {2D} graphene sheets and {1D} nanoribbons, Rep. Prog. Phys. 75
  (2012) 062501.
\newblock \href {https://doi.org/https://doi.org/10.1088/0034-4885/75/6/062501}
  {\path{doi:https://doi.org/10.1088/0034-4885/75/6/062501}}.

\bibitem{Skowron2015}
S.~T. Skowron, I.~V. Lebedeva, A.~M. Popov, E.~Bichoutskaia, Energetics of
  atomic scale structure changes in graphene, Chem. Soc. Rev. 44 (2015)
  3143--3176.
\newblock \href {https://doi.org/https://doi.org/10.1039/C4CS00499J}
  {\path{doi:https://doi.org/10.1039/C4CS00499J}}.

\bibitem{Bhatt2022}
M.~D. Bhatt, H.~Kim, G.~Kim, Various defects in graphene: a review, RSC Adv. 12
  (2022) 21520--21547.
\newblock \href {https://doi.org/https://doi.org/10.1039/D2RA01436J}
  {\path{doi:https://doi.org/10.1039/D2RA01436J}}.

\bibitem{Han2007}
M.~Y. Han, B.~\"Ozyilmaz, Y.~Zhang, P.~Kim, Energy band-gap engineering of
  graphene nanoribbons, Phys. Rev. Lett. 98 (2007) 206805.
\newblock \href {https://doi.org/https://doi.org/10.1103/PhysRevLett.98.206805}
  {\path{doi:https://doi.org/10.1103/PhysRevLett.98.206805}}.

\bibitem{Ritter2009}
K.~A. Ritter, J.~W. Lyding, The influence of edge structure on the electronic
  properties of graphene quantum dots and nanoribbons, Nature Materials 8
  (2009) 235--242.
\newblock \href {https://doi.org/https://doi.org/10.1038/nmat2378}
  {\path{doi:https://doi.org/10.1038/nmat2378}}.

\bibitem{Son2006}
Y.-W. Son, M.~L. Cohen, S.~G. Louie, Energy gaps in graphene nanoribbons, Phys.
  Rev. Lett. 97 (2006) 216803.
\newblock \href {https://doi.org/https://doi.org/10.1103/PhysRevLett.97.216803}
  {\path{doi:https://doi.org/10.1103/PhysRevLett.97.216803}}.

\bibitem{Fujita1996}
M.~Fujita, K.~Wakabayashi, K.~Nakada, K.~Kusakabe, Peculiar localized state at
  zigzag graphite edge, J. Phys. Soc. Jpn. 65 (1996) 1920--1923.
\newblock \href {https://doi.org/https://doi.org/10.1143/JPSJ.65.1920}
  {\path{doi:https://doi.org/10.1143/JPSJ.65.1920}}.

\bibitem{Kobayashi2005}
Y.~Kobayashi, K.-i. Fukui, T.~Enoki, K.~Kusakabe, Y.~Kaburagi, Observation of
  zigzag and armchair edges of graphite using scanning tunneling microscopy and
  spectroscopy, Phys. Rev. B 71 (2005) 193406.
\newblock \href {https://doi.org/https://doi.org/10.1103/PhysRevB.71.193406}
  {\path{doi:https://doi.org/10.1103/PhysRevB.71.193406}}.

\bibitem{Niimi2006}
Y.~Niimi, T.~Matsui, H.~Kambara, K.~Tagami, M.~Tsukada, H.~Fukuyama, Scanning
  tunneling microscopy and spectroscopy of the electronic local density of
  states of graphite surfaces near monoatomic step edges, Phys. Rev. B 73
  (2006) 085421.
\newblock \href {https://doi.org/https://doi.org/10.1103/PhysRevB.73.085421}
  {\path{doi:https://doi.org/10.1103/PhysRevB.73.085421}}.

\bibitem{Tao2011}
C.~Tao, L.~Jiao, O.~V. Yazyev, Y.-C. Chen, J.~Feng, X.~Zhang, R.~B. Capaz,
  J.~M. Tour, A.~Zettl, S.~G. Louie, H.~Dai, M.~F. Crommie, Spatially resolving
  edge states of chiral graphene nanoribbons, Nature Physics 7 (2011) 616--620.
\newblock \href {https://doi.org/https://doi.org/10.1038/nphys1991}
  {\path{doi:https://doi.org/10.1038/nphys1991}}.

\bibitem{Koskinen2008}
P.~Koskinen, S.~Malola, H.~H\"akkinen, Self-passivating edge reconstructions of
  graphene, Phys. Rev. Lett. 101 (2008) 115502.
\newblock \href
  {https://doi.org/https://doi.org/10.1103/PhysRevLett.101.115502}
  {\path{doi:https://doi.org/10.1103/PhysRevLett.101.115502}}.

\bibitem{Li2010}
J.~Li, Z.~Li, G.~Zhou, Z.~Liu, J.~Wu, B.-L. Gu, J.~Ihm, W.~Duan, Spontaneous
  edge-defect formation and defect-induced conductance suppression in graphene
  nanoribbons, Phys. Rev. B 82 (2010) 115410.
\newblock \href {https://doi.org/https://doi.org/10.1103/PhysRevB.82.115410}
  {\path{doi:https://doi.org/10.1103/PhysRevB.82.115410}}.

\bibitem{Gunlycke2007}
D.~Gunlycke, J.~Li, J.~W. Mintmire, C.~T. White, Altering low-bias transport in
  zigzag-edge graphene nanostrips with edge chemistry, Appl. Phys. Lett. 91
  (2007) 112108.
\newblock \href {https://doi.org/https://doi.org/10.1063/1.2783196}
  {\path{doi:https://doi.org/10.1063/1.2783196}}.

\bibitem{Bhandary2010}
S.~Bhandary, O.~Eriksson, B.~Sanyal, M.~I. Katsnelson, Complex edge effects in
  zigzag graphene nanoribbons due to hydrogen loading, Phys. Rev. B 82 (2010)
  165405.
\newblock \href {https://doi.org/https://doi.org/10.1103/PhysRevB.82.165405}
  {\path{doi:https://doi.org/10.1103/PhysRevB.82.165405}}.

\bibitem{Lee2009}
G.~Lee, K.~Cho, Electronic structures of zigzag graphene nanoribbons with edge
  hydrogenation and oxidation, Phys. Rev. B 79 (2009) 165440.
\newblock \href {https://doi.org/https://doi.org/10.1103/PhysRevB.79.165440}
  {\path{doi:https://doi.org/10.1103/PhysRevB.79.165440}}.

\bibitem{Kunstmann2011}
J.~Kunstmann, C.~\"Ozdo\ifmmode~\breve{g}\else \u{g}\fi{}an, A.~Quandt,
  H.~Fehske, Stability of edge states and edge magnetism in graphene
  nanoribbons, Phys. Rev. B 83 (2011) 045414.
\newblock \href {https://doi.org/https://doi.org/10.1103/PhysRevB.83.045414}
  {\path{doi:https://doi.org/10.1103/PhysRevB.83.045414}}.

\bibitem{Nakada1996}
K.~Nakada, M.~Fujita, G.~Dresselhaus, M.~S. Dresselhaus, Edge state in graphene
  ribbons: {Nanometer} size effect and edge shape dependence, Phys. Rev. B 54
  (1996) 17954--17961.
\newblock \href {https://doi.org/https://doi.org/10.1103/PhysRevB.54.17954}
  {\path{doi:https://doi.org/10.1103/PhysRevB.54.17954}}.

\bibitem{Lebedeva2012a}
I.~V. Lebedeva, A.~M. Popov, A.~A. Knizhnik, A.~N. Khlobystov, B.~V. Potapkin,
  Chiral graphene nanoribbon inside a carbon nanotube: \textit{ab initio}
  study, Nanoscale 4 (2012) 4522--4529.
\newblock \href {https://doi.org/https://doi.org/10.1039/C2NR30144J}
  {\path{doi:https://doi.org/10.1039/C2NR30144J}}.

\bibitem{Cantele2009}
G.~Cantele, Y.-S. Lee, D.~Ninno, N.~Marzari, Spin channels in functionalized
  graphene nanoribbons, Nano Letters 9 (2009) 3425--3429.
\newblock \href {https://doi.org/https://doi.org/10.1021/nl901557x}
  {\path{doi:https://doi.org/10.1021/nl901557x}}.

\bibitem{Son2006a}
Y.-W. Son, M.~L. Cohen, S.~G. Louie, Half-metallic graphene nanoribbons, Nature
  444 (2006) 347--349.
\newblock \href {https://doi.org/https://doi.org/10.1038/nature05180}
  {\path{doi:https://doi.org/10.1038/nature05180}}.

\bibitem{Hod2007}
O.~Hod, V.~Barone, J.~E. Peralta, G.~E. Scuseria, Enhanced half-metallicity in
  edge-oxidized zigzag graphene nanoribbons, Nano Letters 7 (2007) 2295--2299.
\newblock \href {https://doi.org/https://doi.org/10.1021/nl0708922}
  {\path{doi:https://doi.org/10.1021/nl0708922}}.

\bibitem{Seitsonen2010}
A.~P. Seitsonen, A.~M. Saitta, T.~Wassmann, M.~Lazzeri, F.~Mauri, Structure and
  stability of graphene nanoribbons in oxygen, carbon dioxide, water, and
  ammonia, Phys. Rev. B 82 (2010) 115425.
\newblock \href {https://doi.org/https://doi.org/10.1103/PhysRevB.82.115425}
  {\path{doi:https://doi.org/10.1103/PhysRevB.82.115425}}.

\bibitem{Magda2014}
G.~Z. Magda, X.~Jin, I.~Hagym{\'a}si, P.~Vancs{\'o}, Z.~Osv{\'a}th,
  P.~Nemes-Incze, C.~Hwang, L.~P. Bir{\'o}, L.~Tapaszt{\'o}, Room-temperature
  magnetic order on zigzag edges of narrow graphene nanoribbons, Nature 514
  (2014) 608--611.
\newblock \href {https://doi.org/https://doi.org/10.1038/nature13831}
  {\path{doi:https://doi.org/10.1038/nature13831}}.

\bibitem{Cheng2012}
Y.~C. Cheng, Z.~Y. Zhu, U.~Schwingenschl{\"o}gl, Mechanical failure of zigzag
  graphene nanoribbons under tensile strain induced by edge reconstruction, J.
  Mater. Chem. 22 (2012) 24676--24680.
\newblock \href {https://doi.org/https://doi.org/10.1039/C2JM34068B}
  {\path{doi:https://doi.org/10.1039/C2JM34068B}}.

\bibitem{Cheng2012prb}
Y.~C. Cheng, H.~T. Wang, Z.~Y. Zhu, Y.~H. Zhu, Y.~Han, X.~X. Zhang,
  U.~Schwingenschl{\"o}gl, Strain-activated edge reconstruction of graphene
  nanoribbons, Phys. Rev. B 85 (2012) 073406.
\newblock \href {https://doi.org/https://doi.org/10.1103/PhysRevB.85.073406}
  {\path{doi:https://doi.org/10.1103/PhysRevB.85.073406}}.

\bibitem{Gan2010}
C.~K. Gan, D.~J. Srolovitz, First-principles study of graphene edge properties
  and flake shapes, Phys. Rev. B 81 (2010) 125445.
\newblock \href {https://doi.org/https://doi.org/10.1103/PhysRevB.81.125445}
  {\path{doi:https://doi.org/10.1103/PhysRevB.81.125445}}.

\bibitem{Song2010}
L.~L. Song, X.~H. Zheng, R.~L. Wang, Z.~Zeng, Dangling bond states, edge
  magnetism, and edge reconstruction in pristine and {B/N}-terminated zigzag
  graphene nanoribbons, J. Phys. Chem. C 114 (2010) 12145--12150.
\newblock \href {https://doi.org/https://doi.org/10.1021/jp1040025}
  {\path{doi:https://doi.org/10.1021/jp1040025}}.

\bibitem{Wang2016}
S.~Wang, L.~Talirz, C.~A. Pignedoli, X.~Feng, K.~M{\"u}llen, R.~Fasel,
  P.~Ruffieux, Giant edge state splitting at atomically precise graphene zigzag
  edges, Nature Communications 7 (2016) 11507.
\newblock \href {https://doi.org/https://doi.org/10.1038/ncomms11507}
  {\path{doi:https://doi.org/10.1038/ncomms11507}}.

\bibitem{Brede2023}
J.~Brede, N.~Merino-D{\'i}ez, A.~Berdonces-Layunta, S.~Sanz,
  A.~Dom{\'i}nguez-Celorrio, J.~Lobo-Checa, M.~Vilas-Varela, D.~Pe{\~{n}}a,
  T.~Frederiksen, J.~I. Pascual, D.~G. de~Oteyza, D.~Serrate, Detecting the
  spin-polarization of edge states in graphene nanoribbons, Nature
  Communications 14 (2023) 6677.
\newblock \href {https://doi.org/https://doi.org/10.1038/s41467-023-42436-7}
  {\path{doi:https://doi.org/10.1038/s41467-023-42436-7}}.

\bibitem{Palacios2010}
J.~J. Palacios, J.~Fern\'andez-Rossier, L.~Brey, H.~A. Fertig, Electronic and
  magnetic structure of graphene nanoribbons, Semiconductor Science and
  Technology 25 (2010) 033003.
\newblock \href {https://doi.org/https://doi.org/10.1088/0268-1242/25/3/033003}
  {\path{doi:https://doi.org/10.1088/0268-1242/25/3/033003}}.

\bibitem{Zhang2014}
W.~Zhang, Voltage-driven spintronic logic gates in graphene nanoribbons,
  Scientific Reports 4 (2014) 6320.
\newblock \href {https://doi.org/https://doi.org/10.1038/srep06320}
  {\path{doi:https://doi.org/10.1038/srep06320}}.

\bibitem{Zeng2011}
M.~Zeng, L.~Shen, M.~Zhou, C.~Zhang, Y.~Feng, Graphene-based bipolar spin diode
  and spin transistor: Rectification and amplification of spin-polarized
  current, Phys. Rev. B 83 (2011) 115427.
\newblock \href {https://doi.org/https://doi.org/10.1103/PhysRevB.83.115427}
  {\path{doi:https://doi.org/10.1103/PhysRevB.83.115427}}.

\bibitem{Zhang2021a}
L.~Zhang, J.~Zhou, H.~Li, L.~Shen, Y.~P. Feng, {Recent progress and challenges
  in magnetic tunnel junctions with {2D} materials for spintronic
  applications}, Appl. Phys. Rev. 8 (2021) 021308.
\newblock \href {https://doi.org/https://doi.org/10.1063/5.0032538}
  {\path{doi:https://doi.org/10.1063/5.0032538}}.

\bibitem{Kang2017}
D.~Kang, B.~Wang, C.~Xia, H.~Li, Perfect spin filter in a tailored zigzag
  graphene nanoribbon, Nanoscale Research Letters 12 (2017) 357.
\newblock \href {https://doi.org/https://doi.org/10.1186/s11671-017-2132-7}
  {\path{doi:https://doi.org/10.1186/s11671-017-2132-7}}.

\bibitem{Rezapour2020}
M.~R. Rezapour, G.~Lee, K.~S. Kim, A high performance {N-doped} graphene
  nanoribbon based spintronic device applicable with a wide range of adatoms,
  Nanoscale Adv. 2 (2020) 5905--5911.
\newblock \href {https://doi.org/https://doi.org/10.1039/D0NA00652A}
  {\path{doi:https://doi.org/10.1039/D0NA00652A}}.

\bibitem{Han2014}
W.~Han, R.~K. Kawakami, M.~Gmitra, J.~Fabian, Graphene spintronics, Nature
  Nanotechnology 9 (2014) 794--807.
\newblock \href {https://doi.org/https://doi.org/10.1038/nnano.2014.214}
  {\path{doi:https://doi.org/10.1038/nnano.2014.214}}.

\bibitem{Wimmer2008}
M.~Wimmer, {\.I}.~Adagideli, S.~Berber, D.~Tom\'anek, K.~Richter, Spin currents
  in rough graphene nanoribbons: {Universal} fluctuations and spin injection,
  Phys. Rev. Lett. 100 (2008) 177207.
\newblock \href
  {https://doi.org/https://doi.org/10.1103/PhysRevLett.100.177207}
  {\path{doi:https://doi.org/10.1103/PhysRevLett.100.177207}}.

\bibitem{Yazyev2008}
O.~V. Yazyev, M.~I. Katsnelson, Magnetic correlations at graphene edges: Basis
  for novel spintronics devices, Phys. Rev. Lett. 100 (2008) 047209.
\newblock \href
  {https://doi.org/https://doi.org/10.1103/PhysRevLett.100.047209}
  {\path{doi:https://doi.org/10.1103/PhysRevLett.100.047209}}.

\bibitem{Yazyev2010}
O.~V. Yazyev, Emergence of magnetism in graphene materials and nanostructures,
  Rep. Prog. Phys. 73 (2010) 056501.
\newblock \href {https://doi.org/https://doi.org/10.1088/0034-4885/73/5/056501}
  {\path{doi:https://doi.org/10.1088/0034-4885/73/5/056501}}.

\bibitem{MunozRojas2009}
F.~{Mu\~noz-Rojas}, J.~Fern\'andez-Rossier, J.~J. Palacios, Giant
  magnetoresistance in ultrasmall graphene based devices, Phys. Rev. Lett. 102
  (2009) 136810.
\newblock \href
  {https://doi.org/https://doi.org/10.1103/PhysRevLett.102.136810}
  {\path{doi:https://doi.org/10.1103/PhysRevLett.102.136810}}.

\bibitem{Kim2008}
W.~Y. Kim, K.~S. Kim, Prediction of very large values of magnetoresistance in a
  graphene nanoribbon device, Nature Nanotechnology 3 (2008) 408--412.
\newblock \href {https://doi.org/https://doi.org/10.1038/nnano.2008.163}
  {\path{doi:https://doi.org/10.1038/nnano.2008.163}}.

\bibitem{Zhang2017}
D.-B. Zhang, S.-H. Wei, Inhomogeneous strain-induced half-metallicity in bent
  zigzag graphene nanoribbons, npj Computational Materials 3 (2017) 32.
\newblock \href {https://doi.org/https://doi.org/10.1038/s41524-017-0036-9}
  {\path{doi:https://doi.org/10.1038/s41524-017-0036-9}}.

\bibitem{Soriano2010}
D.~Soriano, F.~Mu\~noz Rojas, J.~Fern\'andez-Rossier, J.~J. Palacios,
  Hydrogenated graphene nanoribbons for spintronics, Phys. Rev. B 81 (2010)
  165409.
\newblock \href {https://doi.org/https://doi.org/10.1103/PhysRevB.81.165409}
  {\path{doi:https://doi.org/10.1103/PhysRevB.81.165409}}.

\bibitem{Ravagnan2009}
L.~Ravagnan, N.~Manini, E.~Cinquanta, G.~Onida, D.~Sangalli, C.~Motta,
  M.~Devetta, A.~Bordoni, P.~Piseri, P.~Milani, Effect of axial torsion on $sp$
  carbon atomic wires, Phys. Rev. Lett. 102 (2009) 245502.
\newblock \href
  {https://doi.org/https://doi.org/10.1103/PhysRevLett.102.245502}
  {\path{doi:https://doi.org/10.1103/PhysRevLett.102.245502}}.

\bibitem{Poklonski2013}
N.~A. Poklonski, A.~T. Vlassov, S.~A. Vyrko, E.~F. Kislyakov, S.~V. Ratkevich,
  A.~I. Siahlo, Induction: Soliton-like motion of one electron in
  one-dimensional wire with inductance of environment, Word Scientific, 2013,
  pp. 36--39.
\newblock \href {https://doi.org/https://doi.org/10.1142/9789814460187_0007}
  {\path{doi:https://doi.org/10.1142/9789814460187_0007}}.

\bibitem{Zanolli2010}
Z.~Zanolli, G.~Onida, J.-C. Charlier, Quantum spin transport in carbon chains,
  ACS Nano 4 (2010) 5174--5180.
\newblock \href {https://doi.org/https://doi.org/10.1021/nn100712q}
  {\path{doi:https://doi.org/10.1021/nn100712q}}.

\bibitem{Liu2013}
M.~Liu, V.~I. Artyukhov, H.~Lee, F.~Xu, B.~I. Yakobson, Carbyne from first
  principles: Chain of {C} atoms, a nanorod or a nanorope, ACS Nano 7 (2013)
  10075--10082.
\newblock \href {https://doi.org/https://doi.org/10.1021/nn404177r}
  {\path{doi:https://doi.org/10.1021/nn404177r}}.

\bibitem{Artyukhov2014}
V.~I. Artyukhov, M.~Liu, B.~I. Yakobson, Mechanically induced metal–insulator
  transition in carbyne, Nano Letters 14 (2014) 4224--4229.
\newblock \href {https://doi.org/https://doi.org/10.1021/nl5017317}
  {\path{doi:https://doi.org/10.1021/nl5017317}}.

\bibitem{LaTorre2015}
A.~La~Torre, A.~Botello-Mendez, W.~Baaziz, J.-C. Charlier, F.~Banhart,
  Strain-induced metal--semiconductor transition observed in atomic carbon
  chains, Nature Communications 6 (2015) 6636.
\newblock \href {https://doi.org/https://doi.org/10.1038/ncomms7636}
  {\path{doi:https://doi.org/10.1038/ncomms7636}}.

\bibitem{Borrnert2010}
F.~B\"orrnert, C.~B\"orrnert, S.~Gorantla, X.~Liu, A.~Bachmatiuk, J.-O. Joswig,
  F.~R. Wagner, F.~Sch\"affel, J.~H. Warner, R.~Sch\"onfelder, B.~Rellinghaus,
  T.~Gemming, J.~Thomas, M.~Knupfer, B.~B\"uchner, M.~H. R\"ummeli,
  Single-wall-carbon-nanotube/single-carbon-chain molecular junctions, Phys.
  Rev. B 81 (2010) 085439.
\newblock \href {https://doi.org/https://doi.org/10.1103/PhysRevB.81.085439}
  {\path{doi:https://doi.org/10.1103/PhysRevB.81.085439}}.

\bibitem{Crljen2007}
{\v{Z}}.~Crljen, G.~Baranovi\'{c}, Unusual conductance of polyyne-based
  molecular wires, Phys. Rev. Lett. 98 (2007) 116801.
\newblock \href {https://doi.org/https://doi.org/10.1103/PhysRevLett.98.116801}
  {\path{doi:https://doi.org/10.1103/PhysRevLett.98.116801}}.

\bibitem{Lang1998}
N.~D. Lang, P.~Avouris, Oscillatory conductance of carbon-atom wires, Phys.
  Rev. Lett. 81 (1998) 3515--3518.
\newblock \href {https://doi.org/https://doi.org/10.1103/PhysRevLett.81.3515}
  {\path{doi:https://doi.org/10.1103/PhysRevLett.81.3515}}.

\bibitem{Zeng2010}
M.~G. Zeng, L.~Shen, Y.~Q. Cai, Z.~D. Sha, Y.~P. Feng, {Perfect spin-filter and
  spin-valve in carbon atomic chains}, Appl. Phys. Lett. 96 (2010) 042104.
\newblock \href {https://doi.org/https://doi.org/10.1063/1.3299264}
  {\path{doi:https://doi.org/10.1063/1.3299264}}.

\bibitem{Zhou2017}
Y.~Zhou, Y.~Li, J.~Li, J.~Dong, H.~Li, Electronic transport properties of
  carbon and boron nitride chain heterojunctions, J. Mater. Chem. C 5 (2017)
  1165--1178.
\newblock \href {https://doi.org/https://doi.org/10.1039/C6TC04936B}
  {\path{doi:https://doi.org/10.1039/C6TC04936B}}.

\bibitem{Liang2019}
Z.~Liang, X.~Xu, Y.~Jiang, W.~Li, Q.~Wang, G.~Zhang, W.~Q. Tian, Y.~Jiang, The
  influence of coupling between chains on the conductivity of atomic carbon
  chains, Phys. Lett. A 383~(20) (2019) 2409--2415.
\newblock \href
  {https://doi.org/https://doi.org/10.1016/j.physleta.2019.04.053}
  {\path{doi:https://doi.org/10.1016/j.physleta.2019.04.053}}.

\bibitem{Cretu2013}
O.~Cretu, A.~R. Botello-Mendez, I.~Janowska, C.~Pham-Huu, J.-C. Charlier,
  F.~Banhart, Electrical transport measured in atomic carbon chains, Nano
  Letters 13 (2013) 3487--3493.
\newblock \href {https://doi.org/https://doi.org/10.1021/nl4018918}
  {\path{doi:https://doi.org/10.1021/nl4018918}}.

\bibitem{Chen2009}
W.~Chen, A.~V. Andreev, G.~F. Bertsch, Conductance of a single-atom carbon
  chain with graphene leads, Phys. Rev. B 80 (2009) 085410.
\newblock \href {https://doi.org/https://doi.org/10.1103/PhysRevB.80.085410}
  {\path{doi:https://doi.org/10.1103/PhysRevB.80.085410}}.

\bibitem{Liu2015}
X.~Liu, G.~Zhang, Y.-W. Zhang, Tunable mechanical and thermal properties of
  one-dimensional carbyne chain: Phase transition and microscopic dynamics, J.
  Phys. Chem. C 119 (2015) 24156--24164.
\newblock \href {https://doi.org/https://doi.org/10.1021/acs.jpcc.5b08026}
  {\path{doi:https://doi.org/10.1021/acs.jpcc.5b08026}}.

\bibitem{Cahangirov2010}
S.~Cahangirov, M.~Topsakal, S.~Ciraci, Long-range interactions in carbon atomic
  chains, Phys. Rev. B 82 (2010) 195444.
\newblock \href {https://doi.org/https://doi.org/10.1103/PhysRevB.82.195444}
  {\path{doi:https://doi.org/10.1103/PhysRevB.82.195444}}.

\bibitem{Poklonski2019}
N.~A. Poklonski, S.~A. Vyrko, A.~I. Siahlo, O.~N. Poklonskaya, S.~V. Ratkevich,
  N.~N. Hieu, A.~A. Kocherzhenko, Synergy of physical properties of
  low-dimensional carbon-based systems for nanoscale device design, Materials
  Research Express 6 (2019) 042002.
\newblock \href {https://doi.org/https://doi.org/10.1088/2053-1591/aafb1c}
  {\path{doi:https://doi.org/10.1088/2053-1591/aafb1c}}.

\bibitem{Xia2017}
K.~Xia, H.~Zhan, Y.~Gu, Graphene and carbon nanotube hybrid structure: A
  review, Procedia IUTAM 21 (2017) 94--101, 2016 IUTAM Symposium on Nanoscale
  Physical Mechanics.
\newblock \href {https://doi.org/https://doi.org/10.1016/j.piutam.2017.03.042}
  {\path{doi:https://doi.org/10.1016/j.piutam.2017.03.042}}.

\bibitem{Dang2016}
V.~T. Dang, D.~D. Nguyen, T.~T. Cao, P.~H. Le, D.~L. Tran, N.~M. Phan, V.~C.
  Nguyen, Recent trends in preparation and application of carbon
  nanotube–graphene hybrid thin films, Adv. Nat. Sci: Nanosci. Nanotechnol. 7
  (2016) 033002.
\newblock \href {https://doi.org/https://doi.org/10.1088/2043-6262/7/3/033002}
  {\path{doi:https://doi.org/10.1088/2043-6262/7/3/033002}}.

\bibitem{Vinayan2012}
B.~P. Vinayan, R.~Nagar, V.~Raman, N.~Rajalakshmi, K.~S. Dhathathreyan,
  S.~Ramaprabhu, Synthesis of graphene-multiwalled carbon nanotubes hybrid
  nanostructure by strengthened electrostatic interaction and its lithium ion
  battery application, J. Mater. Chem. 22 (2012) 9949--9956.
\newblock \href {https://doi.org/https://doi.org/10.1039/C2JM16294F}
  {\path{doi:https://doi.org/10.1039/C2JM16294F}}.

\bibitem{Laurila2017}
T.~Laurila, S.~Sainio, M.~A. Caro, Hybrid carbon based nanomaterials for
  electrochemical detection of biomolecules, Progress in Materials Science 88
  (2017) 499--594.
\newblock \href {https://doi.org/https://doi.org/10.1016/j.pmatsci.2017.04.012}
  {\path{doi:https://doi.org/10.1016/j.pmatsci.2017.04.012}}.

\bibitem{Navrotskaya2020}
A.~G. Navrotskaya, D.~D. Aleksandrova, E.~F. Krivoshapkina,
  M.~Sillanp\"{a}\"{a}, P.~V. Krivoshapkin, Hybrid materials based on carbon
  nanotubes and nanofibers for environmental applications, Frontiers in
  Chemistry 8 (2020).
\newblock \href {https://doi.org/https://doi.org/10.3389/fchem.2020.00546}
  {\path{doi:https://doi.org/10.3389/fchem.2020.00546}}.

\bibitem{Baughman1987}
R.~H. Baughman, H.~Eckhardt, M.~Kertesz, {Structure‐property predictions for
  new planar forms of carbon: Layered phases containing sp2 and sp atoms}, J.
  Chem. Phys. 87 (1987) 6687--6699.
\newblock \href {https://doi.org/10.1063/1.453405}
  {\path{doi:10.1063/1.453405}}.

\bibitem{Casari2016}
C.~S. Casari, M.~Tommasini, R.~R. Tykwinski, A.~Milani,
  \href{http://dx.doi.org/10.1039/C5NR06175J}{Carbon-atom wires: {1-D} systems
  with tunable properties}, Nanoscale 8 (2016) 4414--4435.
\newblock \href {https://doi.org/10.1039/C5NR06175J}
  {\path{doi:10.1039/C5NR06175J}}.
\newline\urlprefix\url{http://dx.doi.org/10.1039/C5NR06175J}

\bibitem{Srinivasu2012}
K.~Srinivasu, S.~K. Ghosh, Graphyne and graphdiyne: Promising materials for
  nanoelectronics and energy storage applications, J. Phys. Chem. C 116 (2012)
  5951--5956.
\newblock \href {https://doi.org/https://doi.org/10.1021/jp212181h}
  {\path{doi:https://doi.org/10.1021/jp212181h}}.

\bibitem{Ivanovskii2013}
A.~Ivanovskii, Graphynes and graphdyines, Progress in Solid State Chemistry 41
  (2013) 1--19.
\newblock \href
  {https://doi.org/https://doi.org/10.1016/j.progsolidstchem.2012.12.001}
  {\path{doi:https://doi.org/10.1016/j.progsolidstchem.2012.12.001}}.

\bibitem{Li2014}
Y.~Li, L.~Xu, H.~Liu, Y.~Li, Graphdiyne and graphyne: from theoretical
  predictions to practical construction, Chem. Soc. Rev. 43 (2014) 2572--2586.
\newblock \href {https://doi.org/http://dx.doi.org/10.1039/C3CS60388A}
  {\path{doi:http://dx.doi.org/10.1039/C3CS60388A}}.

\bibitem{Song2011}
B.~Song, G.~F. Schneider, Q.~Xu, G.~Pandraud, C.~Dekker, H.~Zandbergen,
  Atomic-scale electron-beam sculpting of near-defect-free graphene
  nanostructures, Nano Letters 11 (2011) 2247--2250.
\newblock \href {https://doi.org/https://doi.org/10.1021/nl200369r}
  {\path{doi:https://doi.org/10.1021/nl200369r}}.

\bibitem{Sinitsa2023}
A.~S. Sinitsa, Y.~G. Polynskaya, I.~V. Lebedeva, A.~A. Knizhnik, A.~M. Popov,
  Precise graphene cutting using a catalyst at a probe tip under an electron
  beam, Phys. Chem. Chem. Phys. 25 (2023) 20715--20727.
\newblock \href {https://doi.org/https://doi.org/10.1039/D3CP00361B}
  {\path{doi:https://doi.org/10.1039/D3CP00361B}}.

\bibitem{Matsokin2023}
N.~A. Matsokin, A.~S. Sinitsa, Y.~G. Polynskaya, I.~V. Lebedeva, A.~A.
  Knizhnik, A.~M. Popov, Formation of carbon propeller-like molecules from
  starphenes under electron irradiation, Phys. Chem. Chem. Phys. 25 (2023)
  27027--27033.
\newblock \href {https://doi.org/https://doi.org/10.1039/D3CP03611A}
  {\path{doi:https://doi.org/10.1039/D3CP03611A}}.

\bibitem{Chuvilin2009}
A.~Chuvilin, J.~C. Meyer, G.~Algara-Siller, U.~Kaiser, From graphene
  constrictions to single carbon chains, New J. Phys. 11 (2009) 083019.
\newblock \href {https://doi.org/https://doi.org/10.1088/1367-2630/11/8/083019}
  {\path{doi:https://doi.org/10.1088/1367-2630/11/8/083019}}.

\bibitem{Chuvilin2010}
A.~Chuvilin, U.~Kaiser, E.~Bichoutskaia, N.~A. Besley, A.~N. Khlobystov, Direct
  transformation of graphene to fullerene, Nature Chemistry 2 (2010) 450--453.
\newblock \href {https://doi.org/https://doi.org/10.1038/nchem.644}
  {\path{doi:https://doi.org/10.1038/nchem.644}}.

\bibitem{Sloan2000}
J.~Sloan, R.~E. Dunin-Borkowski, J.~L. Hutchison, K.~S. Coleman, V.~{Clifford
  Williams}, J.~B. Claridge, A.~P.~E. York, C.~Xu, S.~R. Bailey, G.~Brown,
  S.~Friedrichs, M.~L.~H. Green, The size distribution, imaging and obstructing
  properties of {C$_{60}$} and higher fullerenes formed within arc-grown single
  walled carbon nanotubes, Chem. Phys. Lett. 316 (2000) 191--198.
\newblock \href {https://doi.org/https://doi.org/10.1016/S0009-2614(99)01250-6}
  {\path{doi:https://doi.org/10.1016/S0009-2614(99)01250-6}}.

\bibitem{Sinitsa2018}
A.~S. Sinitsa, I.~V. Lebedeva, A.~M. Popov, A.~A. Knizhnik, Long triple carbon
  chains formation by heat treatment of graphene nanoribbon: {Molecular
  dynamics study with revised Brenner potential}, Carbon 140 (2018) 543--556.
\newblock \href {https://doi.org/https://doi.org/10.1016/j.carbon.2018.08.022}
  {\path{doi:https://doi.org/10.1016/j.carbon.2018.08.022}}.

\bibitem{Sinitsa2017}
A.~S. Sinitsa, T.~W. Chamberlain, T.~Zoberbier, I.~V. Lebedeva, A.~M. Popov,
  A.~A. Knizhnik, R.~L. McSweeney, J.~Biskupek, U.~Kaiser, A.~N. Khlobystov,
  Formation of nickel clusters wrapped in carbon cages: Toward new endohedral
  metallofullerene synthesis, Nano Letters 17 (2017) 1082--1089.
\newblock \href {https://doi.org/https://doi.org/10.1021/acs.nanolett.6b04607}
  {\path{doi:https://doi.org/10.1021/acs.nanolett.6b04607}}.

\bibitem{Jin2009}
C.~Jin, H.~Lan, L.~Peng, K.~Suenaga, S.~Iijima, Deriving carbon atomic chains
  from graphene, Phys. Rev. Lett. 102 (2009) 205501.
\newblock \href {https://doi.org/https://doi.org10.1103/PhysRevLett.102.205501}
  {\path{doi:https://doi.org10.1103/PhysRevLett.102.205501}}.

\bibitem{Sinitsa2021}
A.~S. Sinitsa, I.~V. Lebedeva, Y.~G. Polynskaya, D.~G. de~Oteyza, S.~V.
  Ratkevich, A.~A. Knizhnik, A.~M. Popov, N.~A. Poklonski, Y.~E. Lozovik,
  Transformation of a graphene nanoribbon into a hybrid {1D} nanoobject with
  alternating double chains and polycyclic regions, Phys. Chem. Chem. Phys. 23
  (2021) 425--441.
\newblock \href {https://doi.org/https://doi.org/10.1039/D0CP04090H}
  {\path{doi:https://doi.org/10.1039/D0CP04090H}}.

\bibitem{Tanuma2022}
Y.~Tanuma, P.~Dunk, T.~Maekawa, C.~P. Ewels, Chain formation during hydrogen
  loss and reconstruction in carbon nanobelts, Nanomaterials 12 (2022) 2073.
\newblock \href {https://doi.org/https://doi.org/10.3390/nano12122073}
  {\path{doi:https://doi.org/10.3390/nano12122073}}.

\bibitem{Cai2010}
J.~Cai, P.~Ruffieux, R.~Jaafar, M.~Bieri, T.~Braun, S.~Blankenburg, M.~Muoth,
  A.~P. Seitsonen, M.~Saleh, X.~Feng, K.~M{\"u}llen, R.~Fasel, Atomically
  precise bottom-up fabrication of graphene nanoribbons, Nature 466 (2010)
  470--473.
\newblock \href {https://doi.org/https://doi.org/10.1038/nature09211}
  {\path{doi:https://doi.org/10.1038/nature09211}}.

\bibitem{Zhou2020}
X.~Zhou, G.~Yu, Modified engineering of graphene nanoribbons prepared via
  on-surface synthesis, Advanced Materials 32 (2020) 1905957.
\newblock \href {https://doi.org/https://doi.org/10.1002/adma.201905957}
  {\path{doi:https://doi.org/10.1002/adma.201905957}}.

\bibitem{Lieske2023}
L.-A. Lieske, M.~Commodo, J.~W. Martin, K.~Kaiser, V.~Benekou, P.~Minutolo,
  A.~D’Anna, L.~Gross, Portraits of soot molecules reveal pathways to large
  aromatics, five-/seven-membered rings, and inception through $\pi$-radical
  localization, ACS Nano 17 (2023) 13563--13574.
\newblock \href {https://doi.org/https://doi.org/10.1021/acsnano.3c02194}
  {\path{doi:https://doi.org/10.1021/acsnano.3c02194}}.

\bibitem{Martin2019}
J.~W. Martin, K.~Bowal, A.~Menon, R.~I. Slavchov, J.~Akroyd, S.~Mosbach,
  M.~Kraft, Polar curved polycyclic aromatic hydrocarbons in soot formation,
  Proceedings of the Combustion Institute 37 (2019) 1117--1123.
\newblock \href {https://doi.org/https://doi.org/10.1016/j.proci.2018.05.046}
  {\path{doi:https://doi.org/10.1016/j.proci.2018.05.046}}.

\bibitem{Desyatkin2022}
V.~G. Desyatkin, W.~B. Martin, A.~E. Aliev, N.~E. Chapman, A.~F. Fonseca, D.~S.
  Galv\~{a}o, E.~R. Miller, K.~H. Stone, Z.~Wang, D.~Zakhidov, F.~T. Limpoco,
  S.~R. Almahdali, S.~M. Parker, R.~H. Baughman, V.~O. Rodionov, Scalable
  synthesis and characterization of multilayer $\gamma$-graphyne, new carbon
  crystals with a small direct band gap, Journal of the American Chemical
  Society 144 (2022) 17999--18008.
\newblock \href {https://doi.org/https://doi.org/10.1021/jacs.2c06583}
  {\path{doi:https://doi.org/10.1021/jacs.2c06583}}.

\bibitem{He2023}
S.~He, B.~Wu, Z.~Xia, P.~Guo, Y.~Li, S.~Song, One-pot synthesis of
  gamma-graphyne supported {Pd} nanoparticles with high catalytic activity,
  Nanoscale Adv. 5 (2023) 2487--2492.
\newblock \href {https://doi.org/https://doi.org/10.1039/D3NA00096F}
  {\path{doi:https://doi.org/10.1039/D3NA00096F}}.

\bibitem{Chernozatonskii2009}
L.~A. Chernozatonskii, P.~B. Sorokin, A.~G. Kvashnin, D.~G. Kvashnin,
  Diamond-like {C2H} nanolayer, diamane: Simulation of the structure and
  properties, JETP Letters 90 (2009) 134--138.
\newblock \href {https://doi.org/https://doi.org/10.1134/S0021364009140112}
  {\path{doi:https://doi.org/10.1134/S0021364009140112}}.

\bibitem{Bakharev2020}
P.~V. Bakharev, M.~Huang, M.~Saxena, S.~W. Lee, S.~H. Joo, S.~O. Park, J.~Dong,
  D.~C. Camacho-Mojica, S.~Jin, Y.~Kwon, M.~Biswal, F.~Ding, S.~K. Kwak,
  Z.~Lee, R.~S. Ruoff, Chemically induced transformation of chemical vapour
  deposition grown bilayer graphene into fluorinated single-layer diamond,
  Nature Nanotechnology 15 (2020) 59--66.
\newblock \href {https://doi.org/https://doi.org/10.1038/s41565-019-0582-z}
  {\path{doi:https://doi.org/10.1038/s41565-019-0582-z}}.

\bibitem{Popov2011}
A.~M. Popov, I.~V. Lebedeva, A.~A. Knizhnik, Y.~E. Lozovik, B.~V. Potapkin,
  Commensurate-incommensurate phase transition in bilayer graphene, Phys. Rev.
  B 84 (2011) 045404.
\newblock \href {https://doi.org/https://doi.org/10.1103/PhysRevB.84.045404}
  {\path{doi:https://doi.org/10.1103/PhysRevB.84.045404}}.

\bibitem{Alden2013}
J.~S. Alden, A.~W. Tsen, P.~Y. Huang, R.~Hovden, L.~Brown, J.~Park, D.~A.
  Muller, P.~L. McEuen, Strain solitons and topological defects in bilayer
  graphene, Proceedings of the National Academy of Sciences 110 (2013)
  11256--11260.
\newblock \href {https://doi.org/https://doi.org/10.1073/pnas.1309394110}
  {\path{doi:https://doi.org/10.1073/pnas.1309394110}}.

\bibitem{Li2023}
Z.~Li, J.~Han, S.~Cao, Z.~Zhang, X.~Deng, Physical properties of monolayer
  $\mathrm{Mn}{(\mathrm{BiTeS})}_{2}$ and its applications in sub--3 nm
  spintronic devices, Phys. Rev. B 108 (2023) 184413.
\newblock \href {https://doi.org/https://doi.org/10.1103/PhysRevB.108.184413}
  {\path{doi:https://doi.org/10.1103/PhysRevB.108.184413}}.

\bibitem{Davies2021}
F.~H. Davies, C.~J. Price, N.~T. Taylor, S.~G. Davies, S.~P. Hepplestone, Band
  alignment of transition metal dichalcogenide heterostructures, Phys. Rev. B
  103 (2021) 045417.
\newblock \href {https://doi.org/https://doi.org/10.1103/PhysRevB.103.045417}
  {\path{doi:https://doi.org/10.1103/PhysRevB.103.045417}}.

\bibitem{Gopalan2022}
S.~Gopalan, M.~L. Van~de Put, G.~Gaddemane, M.~V. Fischetti, Theoretical study
  of electronic transport in two-dimensional transition metal dichalcogenides:
  Effects of the dielectric environment, Phys. Rev. Appl. 18 (2022) 054062.
\newblock \href
  {https://doi.org/https://doi.org/10.1103/PhysRevApplied.18.054062}
  {\path{doi:https://doi.org/10.1103/PhysRevApplied.18.054062}}.

\bibitem{Dias2021}
A.~C. Dias, H.~Bragan\c{c}a, J.~P.~A. de~Mendon\c{c}a, J.~L.~F. Da~Silva,
  Excitonic effects on two-dimensional transition-metal dichalcogenide
  monolayers: Impact on solar cell efficiency, ACS Applied Energy Materials 4
  (2021) 3265--3278.
\newblock \href {https://doi.org/https://doi.org/10.1021/acsaem.0c03039}
  {\path{doi:https://doi.org/10.1021/acsaem.0c03039}}.

\bibitem{Lebedeva2008}
I.~V. Lebedeva, A.~A. Knizhnik, A.~A. Bagatur’yants, B.~V. Potapkin, Kinetics
  of {2D--3D} transformations of carbon nanostructures, Physica E:
  Low-dimensional Systems and Nanostructures 40 (2008) 2589--2595.
\newblock \href {https://doi.org/https://doi.org/10.1016/j.physe.2007.09.155}
  {\path{doi:https://doi.org/10.1016/j.physe.2007.09.155}}.

\bibitem{Skowron2013}
S.~T. Skowron, I.~V. Lebedeva, A.~M. Popov, E.~Bichoutskaia, Approaches to
  modelling irradiation-induced processes in transmission electron microscopy,
  Nanoscale 5 (2013) 6677--6692.
\newblock \href {https://doi.org/https://doi.org/10.1039/C3NR02130K}
  {\path{doi:https://doi.org/10.1039/C3NR02130K}}.

\bibitem{He2015}
K.~He, A.~W. Robertson, Y.~Fan, C.~S. Allen, Y.-C. Lin, K.~Suenaga, A.~I.
  Kirkland, J.~H. Warner, Temperature dependence of the reconstruction of
  zigzag edges in graphene, ACS Nano 9 (2015) 4786--4795.
\newblock \href {https://doi.org/https://doi.org/10.1021/acsnano.5b01130}
  {\path{doi:https://doi.org/10.1021/acsnano.5b01130}}.

\bibitem{Polynskaya2022}
Y.~G. Polynskaya, I.~V. Lebedeva, A.~A. Knizhnik, A.~M. Popov, Optimal model of
  semi-infinite graphene for \textit{ab initio} calculations of reactions at
  graphene edges by the example of zigzag edge reconstruction, Computational
  and Theoretical Chemistry 1214 (2022) 113755.
\newblock \href {https://doi.org/https://doi.org/10.1016/j.comptc.2022.113755}
  {\path{doi:https://doi.org/10.1016/j.comptc.2022.113755}}.

\bibitem{Polynskaya2022a}
Y.~G. Polynskaya, I.~V. Lebedeva, A.~A. Knizhnik, A.~M. Popov, Reconstruction
  of zigzag graphene edges: Energetics, kinetics, and residual defects, J.
  Phys. Chem. Lett. 13 (2022) 10326--10330.
\newblock \href {https://doi.org/https://doi.org/10.1021/acs.jpclett.2c02706}
  {\path{doi:https://doi.org/10.1021/acs.jpclett.2c02706}}.

\bibitem{Wassmann2008}
T.~Wassmann, A.~P. Seitsonen, A.~M. Saitta, M.~Lazzeri, F.~Mauri, Structure,
  stability, edge states, and aromaticity of graphene ribbons, Phys. Rev. Lett.
  101 (2008) 096402.
\newblock \href
  {https://doi.org/https://doi.org/10.1103/PhysRevLett.101.096402}
  {\path{doi:https://doi.org/10.1103/PhysRevLett.101.096402}}.

\bibitem{Tommasini2014}
M.~Tommasini, A.~Milani, D.~Fazzi, A.~Lucotti, C.~Castiglioni, J.~A.
  Januszewski, D.~Wendinger, R.~R. Tykwinski, $\pi$-conjugation and end group
  effects in long cumulenes: Raman spectroscopy and {DFT} calculations, J.
  Phys. Chem. C 118 (2014) 26415--26425.
\newblock \href {https://doi.org/https://doi.org/10.1021/jp509724d}
  {\path{doi:https://doi.org/10.1021/jp509724d}}.

\bibitem{Gu2008}
X.~Gu, R.~I. Kaiser, A.~M. Mebel, Chemistry of energetically activated
  cumulenes -- from allene {(H$_2$CCCH$_2$)} to hexapentaene
  {(H$_2$CCCCCCH$_2$)}, Chem. Phys. Chem. 9 (2008) 350--369.
\newblock \href {https://doi.org/https://doi.org/10.1002/cphc.200700609}
  {\path{doi:https://doi.org/10.1002/cphc.200700609}}.

\bibitem{Hino2003}
S.~Hino, Y.~Okada, K.~Iwasaki, M.~Kijima, H.~Shirakawa, Electronic structures
  of cumulene type carbyne model compounds: a typical example of
  one-dimensional quantum well, Chem. Phys. Lett. 372 (2003) 59--65.
\newblock \href {https://doi.org/https://doi.org/10.1016/S0009-2614(03)00360-9}
  {\path{doi:https://doi.org/10.1016/S0009-2614(03)00360-9}}.

\bibitem{Eisler2005}
S.~Eisler, A.~D. Slepkov, E.~Elliott, T.~Luu, R.~McDonald, F.~A. Hegmann, R.~R.
  Tykwinski, Polyynes as a model for carbyne: synthesis, physical properties,
  and nonlinear optical response, J. Am. Chem. Soc. 127 (2005) 2666--2676.
\newblock \href {https://doi.org/https://doi.org/10.1021/ja044526l}
  {\path{doi:https://doi.org/10.1021/ja044526l}}.

\bibitem{Yueze2022}
Y.~Gao, R.~R. Tykwinski, Advances in polyynes to model carbyne, Acc. Chem. Res.
  55 (2022) 3616--3630.
\newblock \href {https://doi.org/https://doi.org/10.1021/acs.accounts.2c00662}
  {\path{doi:https://doi.org/10.1021/acs.accounts.2c00662}}.

\bibitem{Agarwal2016}
N.~R. Agarwal, A.~Lucotti, M.~Tommasini, W.~A. Chalifoux, R.~R. Tykwinski,
  Nonlinear optical properties of polyynes: An experimental prediction for
  carbyne, J. Phys. Chem. C 120 (2016) 11131--11139.
\newblock \href {https://doi.org/https://doi.org/10.1021/acs.jpcc.6b03071}
  {\path{doi:https://doi.org/10.1021/acs.jpcc.6b03071}}.

\bibitem{Kertesz1978}
M.~Kertesz, J.~Koller, A.~A\v{z}man, {Ab initio Hartree-Fock} crystal orbital
  studies. {II}. {Energy} bands of an infinite carbon chain, J. Chem. Phys. 68
  (1978) 2779--2782.
\newblock \href {https://doi.org/https://doi.org/10.1063/1.436070}
  {\path{doi:https://doi.org/10.1063/1.436070}}.

\bibitem{Jain2011}
M.~Jain, J.~R. Chelikowsky, S.~G. Louie, Reliability of hybrid functionals in
  predicting band gaps, Phys. Rev. Lett. 107 (2011) 216806.
\newblock \href
  {https://doi.org/https://doi.org/10.1103/PhysRevLett.107.216806}
  {\path{doi:https://doi.org/10.1103/PhysRevLett.107.216806}}.

\bibitem{Borlido2020}
P.~Borlido, J.~Schmidt, A.~W. Huran, F.~Tran, M.~A.~L. Marques, S.~Botti,
  Exchange-correlation functionals for band gaps of solids: benchmark,
  reparametrization and machine learning, {npj Computational Materials} 6
  (2020) 96.
\newblock \href {https://doi.org/https://doi.org/10.1038/s41524-020-00360-0}
  {\path{doi:https://doi.org/10.1038/s41524-020-00360-0}}.

\bibitem{Heyd2004}
J.~Heyd, G.~E. Scuseria, {Efficient hybrid density functional calculations in
  solids: Assessment of the Heyd–Scuseria–Ernzerhof screened Coulomb hybrid
  functional}, J. Chem. Phys. 121 (2004) 1187--1192.
\newblock \href {https://doi.org/https://doi.org/10.1063/1.1760074}
  {\path{doi:https://doi.org/10.1063/1.1760074}}.

\bibitem{Heyd2005}
J.~Heyd, J.~E. Peralta, G.~E. Scuseria, R.~L. Martin, {Energy band gaps and
  lattice parameters evaluated with the Heyd-Scuseria-Ernzerhof screened hybrid
  functional}, J. Chem. Phys. 123 (2005) 174101.
\newblock \href {https://doi.org/https://doi.org/10.1063/1.2085170}
  {\path{doi:https://doi.org/10.1063/1.2085170}}.

\bibitem{Perdew1996}
J.~P. Perdew, K.~Burke, M.~Ernzerhof, Generalized gradient approximation made
  simple, Phys. Rev. Lett. 77 (1996) 3865--3868.
\newblock \href {https://doi.org/https://doi.org/10.1103/PhysRevLett.77.3865}
  {\path{doi:https://doi.org/10.1103/PhysRevLett.77.3865}}.

\bibitem{Heyd2003}
J.~Heyd, G.~E. Scuseria, M.~Ernzerhof, {Hybrid functionals based on a screened
  Coulomb potential}, J. Chem. Phys. 118 (2003) 8207--8215.
\newblock \href {https://doi.org/https://doi.org/10.1063/1.1564060}
  {\path{doi:https://doi.org/10.1063/1.1564060}}.

\bibitem{Heyd2006}
J.~Heyd, G.~E. Scuseria, M.~Ernzerhof, {Erratum: “Hybrid functionals based on
  a screened Coulomb potential” [J. Chem. Phys. 118, 8207 (2003)]}, J. Chem.
  Phys. 124 (2006) 219906.
\newblock \href {https://doi.org/https://doi.org/10.1063/1.2204597}
  {\path{doi:https://doi.org/10.1063/1.2204597}}.

\bibitem{Barone2005a}
V.~Barone, J.~E. Peralta, G.~E. Scuseria, Optical transitions in metallic
  single-walled carbon nanotubes, Nano Letters 5 (2005) 1830--1833.
\newblock \href {https://doi.org/https://doi.org/10.1021/nl0509733}
  {\path{doi:https://doi.org/10.1021/nl0509733}}.

\bibitem{Barone2005}
V.~Barone, J.~E. Peralta, M.~Wert, J.~Heyd, G.~E. Scuseria, Density functional
  theory study of optical transitions in semiconducting single-walled carbon
  nanotubes, Nano Letters 5 (2005) 1621--1624.
\newblock \href {https://doi.org/https://doi.org/10.1021/nl0506352}
  {\path{doi:https://doi.org/10.1021/nl0506352}}.

\bibitem{Giannozzi2017}
P.~Giannozzi, O.~Andreussi, T.~Brumme, O.~Bunau, M.~B. Nardelli, M.~Calandra,
  R.~Car, C.~Cavazzoni, D.~Ceresoli, M.~Cococcioni, N.~Colonna, I.~Carnimeo,
  A.~D. Corso, S.~de~Gironcoli, P.~Delugas, R.~A. DiStasio, A.~Ferretti,
  A.~Floris, G.~Fratesi, G.~Fugallo, R.~Gebauer, U.~Gerstmann, F.~Giustino,
  T.~Gorni, J.~Jia, M.~Kawamura, H.-Y. Ko, A.~Kokalj,
  E.~K{\"u}\c{c}{\"u}kbenli, M.~Lazzeri, M.~Marsili, N.~Marzari, F.~Mauri,
  N.~L. Nguyen, H.-V. Nguyen, A.~{Otero-de-la-Roza}, L.~Paulatto, S.~Ponc\'{e},
  D.~Rocca, R.~Sabatini, B.~Santra, M.~Schlipf, A.~P. Seitsonen, A.~Smogunov,
  I.~Timrov, T.~Thonhauser, P.~Umari, N.~Vast, X.~Wu, S.~Baroni, Advanced
  capabilities for materials modelling with {Quantum ESPRESSO}, J. Phys.:
  Condens. Matter 29 (2017) 465901.
\newblock \href {https://doi.org/https://doi.org/10.1088/1361-648X/aa8f79}
  {\path{doi:https://doi.org/10.1088/1361-648X/aa8f79}}.

\bibitem{Giannozzi2009}
P.~Giannozzi, S.~Baroni, N.~Bonini, M.~Calandra, R.~Car, C.~Cavazzoni,
  D.~Ceresoli, G.~L. Chiarotti, M.~Cococcioni, I.~Dabo, A.~D. Corso,
  S.~de~Gironcoli, S.~Fabris, G.~Fratesi, R.~Gebauer, U.~Gerstmann,
  C.~Gougoussis, A.~Kokalj, M.~Lazzeri, L.~Martin-Samos, N.~Marzari, F.~Mauri,
  R.~Mazzarello, S.~Paolini, A.~Pasquarello, L.~Paulatto, C.~Sbraccia,
  S.~Scandolo, G.~Sclauzero, A.~P. Seitsonen, A.~Smogunov, P.~Umari, R.~M.
  Wentzcovitch, {QUANTUM ESPRESSO}: a modular and open-source software project
  for quantum simulations of materials, J. Phys.: Condens. Matter 21 (2009)
  395502.
\newblock \href
  {https://doi.org/https://doi.org/10.1088/0953-8984/21/39/395502}
  {\path{doi:https://doi.org/10.1088/0953-8984/21/39/395502}}.

\bibitem{QE}
Quantum ESPRESSO code: http://www.quantum-espresso.org for BLAS technical
  forum, 2020.

\bibitem{Methfesse1989}
M.~Methfessel, A.~T. Paxton, High-precision sampling for {Brillouin-zone}
  integration in metals, Phys. Rev. B 40 (1989) 3616--3621.
\newblock \href {https://doi.org/https://doi.org/10.1103/PhysRevB.40.3616}
  {\path{doi:https://doi.org/10.1103/PhysRevB.40.3616}}.

\bibitem{DalCorso2014}
A.~{Dal Corso}, Pseudopotentials periodic table: {From H to Pu}, Computational
  Materials Science 95 (2014) 337--350.
\newblock \href
  {https://doi.org/https://doi.org/10.1016/j.commatsci.2014.07.043}
  {\path{doi:https://doi.org/10.1016/j.commatsci.2014.07.043}}.

\bibitem{Hartwigsen1998}
C.~Hartwigsen, S.~Goedecker, J.~Hutter, Relativistic separable dual-space
  {Gaussian} pseudopotentials {from H to Rn}, Phys. Rev. B 58 (1998)
  3641--3662.
\newblock \href {https://doi.org/https://doi.org/10.1103/PhysRevB.58.3641}
  {\path{doi:https://doi.org/10.1103/PhysRevB.58.3641}}.

\bibitem{ASAP}
{Simune Atomistics, S.L.}, {ASAP-2024.0 — Atomistic Simulation Advanced
  Platform}, https://www.simuneatomistics.com, accessed: 23.04.2024 (2024).

\bibitem{Momma2011}
K.~Momma, F.~Izumi, {\it VESTA 3} for three-dimensional visualization of
  crystal, volumetric and morphology data, J. Appl. Crystallogr. 44 (2011)
  1272--1276.
\newblock \href {https://doi.org/https://doi.org/10.1107/S0021889811038970}
  {\path{doi:https://doi.org/10.1107/S0021889811038970}}.

\bibitem{Pisani2007}
L.~Pisani, J.~A. Chan, B.~Montanari, N.~M. Harrison, Electronic structure and
  magnetic properties of graphitic ribbons, Phys. Rev. B 75 (2007) 064418.
\newblock \href {https://doi.org/https://doi.org/10.1103/PhysRevB.75.064418}
  {\path{doi:https://doi.org/10.1103/PhysRevB.75.064418}}.

\bibitem{Vyrko2024}
S.~A. Vyrko, Y.~G. Polynskaya, N.~A. Matsokin, A.~M. Popov, A.~A. Knizhnik,
  N.~A. Poklonski, Y.~E. Lozovik, Carbon nanobracelets, Chem. Phys. Lett. 835
  (2024) 140999.
\newblock \href {https://doi.org/https://doi.org/10.1016/j.cplett.2023.140999}
  {\path{doi:https://doi.org/10.1016/j.cplett.2023.140999}}.

\bibitem{Poklonski2012}
N.~A. Poklonski, S.~V. Ratkevich, S.~A. Vyrko, E.~F. Kislyakov, O.~N. Bubel’,
  A.~M. Popov, Y.~E. Lozovik, N.~N. Hieu, N.~A. Viet, Structural phase
  transition and band gap of uniaxially deformed (6,0) carbon nanotube,
  Chemical Physics Letters 545 (2012) 71--77.
\newblock \href {https://doi.org/https://doi.org/10.1016/j.cplett.2012.07.023}
  {\path{doi:https://doi.org/10.1016/j.cplett.2012.07.023}}.

\bibitem{Methfessel1968}
S.~{Methfessel}, D.~C. {Mattis}, {Magnetic Semiconductors}, Handbuch der Physik
  4 (1968) 389--562.
\newblock \href {https://doi.org/https://doi.org/10.1007/978-3-642-46132-3_5}
  {\path{doi:https://doi.org/10.1007/978-3-642-46132-3_5}}.

\bibitem{Pappas2013}
S.~D. Pappas, P.~Poulopoulos, B.~Lewitz, A.~Straub, A.~Goschew, V.~Kapaklis,
  F.~Wilhelm, A.~Rogalev, P.~Fumagalli, Direct evidence for significant
  spin-polarization of {EuS in Co/EuS} multilayers at room temperature,
  Scientific Reports 3 (2013) 1333.
\newblock \href {https://doi.org/https://doi.org/10.1038/srep01333}
  {\path{doi:https://doi.org/10.1038/srep01333}}.

\bibitem{Ohno1996}
H.~Ohno, A.~Shen, F.~Matsukura, A.~Oiwa, A.~Endo, S.~Katsumoto, Y.~Iye,
  {(Ga,Mn)As: A new diluted magnetic semiconductor based on GaAs}, Appl. Phys.
  Lett. 69 (1996) 363--365.
\newblock \href {https://doi.org/https://doi.org/10.1063/1.118061}
  {\path{doi:https://doi.org/10.1063/1.118061}}.

\bibitem{Kalita2023}
H.~Kalita, M.~Bhushan, L.~{Robindro Singh}, A comprehensive review on
  theoretical concepts, types and applications of magnetic semiconductors,
  Materials Science and Engineering: B 288 (2023) 116201.
\newblock \href {https://doi.org/https://doi.org/10.1016/j.mseb.2022.116201}
  {\path{doi:https://doi.org/10.1016/j.mseb.2022.116201}}.

\bibitem{Cheng2022}
R.~Cheng, L.~Yin, Y.~Wen, B.~Zhai, Y.~Guo, Z.~Zhang, W.~Liao, W.~Xiong,
  H.~Wang, S.~Yuan, J.~Jiang, C.~Liu, J.~He, Ultrathin ferrite nanosheets for
  room-temperature two-dimensional magnetic semiconductors, Nature
  Communications 13 (2022) 5241.
\newblock \href {https://doi.org/https://doi.org/10.1038/s41467-022-33017-1}
  {\path{doi:https://doi.org/10.1038/s41467-022-33017-1}}.

\bibitem{Wolf2001}
S.~A. Wolf, D.~D. Awschalom, R.~A. Buhrman, J.~M. Daughton, S.~von Moln\'ar,
  M.~L. Roukes, A.~Y. Chtchelkanova, D.~M. Treger, Spintronics: {A} spin-based
  electronics vision for the future, Science 294~(5546) (2001) 1488--1495.
\newblock \href {https://doi.org/https://doi.org/10.1126/science.1065389}
  {\path{doi:https://doi.org/10.1126/science.1065389}}.

\bibitem{Kioseoglou2004}
G.~Kioseoglou, A.~T. Hanbicki, J.~M. Sullivan, O.~M.~J. van~'t Erve, C.~H. Li,
  S.~C. Erwin, R.~Mallory, M.~Yasar, A.~Petrou, B.~T. Jonker, Electrical spin
  injection from an n-type ferromagnetic semiconductor into a {III--V} device
  heterostructure, Nature Materials 3 (2004) 799--803.
\newblock \href {https://doi.org/https://doi.org/10.1038/nmat1239}
  {\path{doi:https://doi.org/10.1038/nmat1239}}.

\bibitem{Telegin2022}
A.~Telegin, Y.~Sukhorukov, Magnetic semiconductors as materials for
  spintronics, Magnetochemistry 8~(12) (2022) 173.
\newblock \href
  {https://doi.org/https://doi.org/10.3390/magnetochemistry8120173}
  {\path{doi:https://doi.org/10.3390/magnetochemistry8120173}}.

\bibitem{Zutic2004}
I.~\ifmmode \check{Z}\else \v{Z}\fi{}uti\ifmmode~\acute{c}\else \'{c}\fi{},
  J.~Fabian, S.~Das~Sarma, Spintronics: {Fundamentals} and applications, Rev.
  Mod. Phys. 76 (2004) 323--410.
\newblock \href {https://doi.org/https://doi.org/10.1103/RevModPhys.76.323}
  {\path{doi:https://doi.org/10.1103/RevModPhys.76.323}}.

\bibitem{Peng2014}
S.~Peng, Y.~Zhang, M.~Wang, Y.~Zhang, W.~Zhao, Magnetic Tunnel Junctions for
  Spintronics: {Principles} and Applications, John Wiley \& Sons, Ltd., 2014,
  pp. 1--16.
\newblock \href {https://doi.org/https://doi.org/10.1002/047134608X.W8231}
  {\path{doi:https://doi.org/10.1002/047134608X.W8231}}.

\end{thebibliography}

\end{document}